\documentclass[acmsmall]{acmart}

\AtBeginDocument{%
  }

\usepackage{amsmath,amsfonts}
\usepackage{graphicx}
\usepackage{textcomp}
\usepackage{xcolor}
\usepackage{caption}
\usepackage{lipsum}
\usepackage{makecell}
\usepackage{xspace}
\usepackage{float}
\usepackage{tipa}

\usepackage{arydshln}
\usepackage{enumitem}
\usepackage{multirow}
\usepackage{tcolorbox}
\usepackage{colortbl}
\usepackage[export]{adjustbox} % for \sbox
\usepackage[T1]{fontenc}
\usepackage{tabularx}
\usepackage{threeparttable}
\usepackage{listings}
\usepackage{xcolor}
\usepackage{booktabs}
\usepackage{subcaption}
\usepackage{algorithm}
\usepackage{algorithmic}

\usepackage{changes}

\newcommand{\methodname}{DCGen\xspace}
\definecolor{codegreen}{rgb}{0,0.6,0}
\newcommand{\add}[1]{\textcolor{black}{{#1}}}

\newsavebox{\arrangebox}
\newlength{\arrangeht}
 \newcommand{\advColor}[1]{%
  \pgfmathsetmacro{\val}{#1}%
  \ifdim \val pt < 0pt
    \textcolor{red}{#1}%
  \else
    \ifdim \val pt > 5pt
      \underline{\textcolor{codegreen}{#1}}%
    \else
      \textcolor{codegreen}{#1}%
    \fi
  \fi
}
\setcopyright{cc}
\setcctype{by}
\acmDOI{10.1145/3729364}
\acmYear{2025}
\acmJournal{PACMSE}
\acmVolume{2}
\acmNumber{FSE}
\acmArticle{FSE094}
\acmMonth{7}
\received{2024-09-13}
\received[accepted]{2025-04-01}

\begin{document}

\title{Divide-and-Conquer: Generating UI Code from Screenshots}

\author{Yuxuan Wan}
\orcid{0009-0006-6739-4675}
\affiliation{%
  \institution{The Chinese University of Hong Kong}
  \city{Hong Kong}
  \country{China}
}
\email{yxwan9@cse.cuhk.edu.hk}

\author{Chaozheng Wang}
\orcid{0000-0002-3935-7328}
\affiliation{%
  \institution{The Chinese University of Hong Kong}
  \city{Hong Kong}
  \country{China}
}
\email{adf111178@gmail.com}

\author{Yi Dong}
\orcid{0009-0007-4456-072X}
\affiliation{%
  \institution{The Chinese University of Hong Kong}
  \city{Hong Kong}
  \country{China}
}
\email{1155173962@link.cuhk.edu.hk}

\author{Wenxuan Wang}
\orcid{0000-0002-9803-8204}
\affiliation{%
  \institution{The Chinese University of Hong Kong}
  \city{Hong Kong}
  \country{China}
}
\email{wxwang@cse.cuhk.edu.hk}

\author{Shuqing Li}
\orcid{0000-0001-6323-1402}
\affiliation{%
  \institution{The Chinese University of Hong Kong}
  \city{Hong Kong}
  \country{China}
}
\email{sqli21@cse.cuhk.edu.hk}

\author{Yintong Huo}
\authornote{Yintong Huo is the corresponding author.}
\orcid{0009-0006-8798-5667}
\affiliation{%
  \institution{Singapore Management University}
  \city{Singapore}
  \country{Singapore}
}
\email{ythuo@smu.edu.sg}

\author{Michael Lyu}
\orcid{0000-0002-3666-5798}
\affiliation{%
  \institution{The Chinese University of Hong Kong}
  \city{Hong Kong}
  \country{China}
}
\email{lyu@cse.cuhk.edu.hk}


\begin{abstract}
Websites are critical in today's digital world, with over 1.11 billion currently active and approximately 252,000 new sites launched daily. Converting website layout design into functional UI code is a time-consuming yet indispensable step of website development. Manual methods of converting visual designs into functional code present significant challenges, especially for non-experts. 
To explore automatic design-to-code solutions, we first conduct a motivating study on GPT-4o and identify three types of issues in generating UI code: element omission, element distortion, and element misarrangement.
We further reveal that a focus on smaller visual segments can help multimodal large language models (MLLMs) mitigate these failures in the generation process.

In this paper, we propose DCGen, a divide-and-conquer-based approach to automate the translation of webpage design to UI code. DCGen starts by dividing screenshots into manageable segments, generating code for each segment, and then reassembling them into complete UI code for the entire screenshot. We conduct extensive testing with a dataset comprised of real-world websites and various MLLMs and demonstrate that DCGen achieves up to a 15\% improvement in visual similarity and 8\% in code similarity for large input images. Human evaluations show that DCGen can help developers implement webpages significantly faster and more similar to the UI designs. To the best of our knowledge, DCGen is the first segment-aware MLLM-based approach for generating UI code directly from screenshots.
\end{abstract}


\begin{CCSXML}
<ccs2012>
   <concept>
    <concept_id>10011007.10011074.10011092.10011782</concept_id>
       <concept_desc>Software and its engineering~Automatic programming</concept_desc>
       <concept_significance>500</concept_significance>
       </concept>
   <concept>
       <concept_id>10010147.10010178</concept_id>
       <concept_desc>Computing methodologies~Artificial intelligence</concept_desc>
       <concept_significance>300</concept_significance>
       </concept>
 </ccs2012>
\end{CCSXML}

\ccsdesc[500]{Software and its engineering~Automatic programming}
\ccsdesc[300]{Computing methodologies~Artificial intelligence}

\keywords{Multi-modal Large Language Model, Code Generation, User Interface, Web Development}

\maketitle

\section{Introduction}

%%% Importance of website
Websites are important in today's digital landscape, serving as essential platforms for diverse applications in our daily lives. Statistics show that there are over 1.1 billion online websites, with approximately 252,000 new websites being created every day~\cite{website_statistics_2024, wordpress_statistics_2024}.

%%%% stages in design website
Developing a website's graphical user interface (GUI) involves two sequential stages: UI design and UI implementation. During the design stage, graphic artists create and customize the website layout, which is then provided to developers as pixel-based concept drawings~\cite{Chen2018FromUI, Kuusinen2013DesigningUE}. Subsequently, the implementation stage involves translating these visual designs into functional UI code using GUI frameworks. Screenshots of the GUI can effectively approximate the original UI designs, as they capture the intended pixel-level layout and visual structure of the design before being converted into operational code.

%%% manual approach is inefficient.
Manually implementing visual designs of websites into functional code is complicated and time-consuming, as it requires domain knowledge of mapping GUI framework elements to spatial layouts. For laypersons, this knowledge barrier has hindered them from efficiently building their own web applications, even if they have concrete ideas for the design~\cite{Si2024Design2CodeHF}. 
For developers, acquiring such knowledge is challenging, time-consuming, and error-prone due to the intricate nature of GUI frameworks, marked by the excessive amount of GUI components, the complexity of spatial layouts, and the diversity of GUI frameworks~\cite{Chen2018FromUI, Nguyen2015ReverseEM, Lelli2015ClassifyingAQ, Moran2018AutomatedRO}. Although modern GUI builders provide interactive features (e.g., drag \& drop, what-you-see-is-what-you-get) for developing a GUI, these tools often limit customization~\cite{Nguyen2015ReverseEM}, causing discrepancies between the intended design and actual implementation, and are known to introduce bugs and presentation failures even for simple tasks~\cite{Zeidler2013EvaluatingDM}.

%%% the benefits of automatic approach.
The challenges in GUI production necessitate an automated approach to translating UI designs (e.g., images, Photoshop documents) into GUI code, referred to as \textit{Design-to-Code}. The automated design-to-code techniques offer benefits to individuals and companies. First, it has the potential to democratize front-end web application development~\cite{Nguyen2015ReverseEM}, allowing non-experts to easily build web applications. Second, it allows smaller companies to focus more on visual features, rather than on translating designs into functional code~\cite{Moran2018MachineLP}.

  Despite its promising benefits, practical solutions for generating code implementation from website design have been underexplored. A line of research made efforts to transform GUI design into code for Android mobile apps but these approaches are not generalizable to the website design containing complicated interfaces~\cite{Chen2018FromUI, Nguyen2015ReverseEM, Abdelhamid2020DeepLP, Moran2018MachineLP}. To enable finger-friendly touch targets, mobile apps utilize minimalist designs to avoid clutter on small screens, resulting in fewer visual elements and simpler structures. Additionally, mobile development relies on platform-specific tools~\cite{Jabangwe2018SoftwareEP} (e.g., iOS uses Xcode, Android uses Android Studio), which do not apply to website development that typically uses HTML, CSS, and JavaScript~\cite{bitcot_2024, knowledgehut_2024, inkbotdesign_2024}.
Another line of research generates website UI code from screenshots using deep learning-based techniques (e.g., CNN, LSTM), but they rely on synthetic data. 
For example, Beltramelli et al.~\cite{Beltramelli2018Pix2c} integrates CNN and RNN models to generate code from screenshots. However, this method is limited to a narrow range of simple user interface designs, accommodating only five types of web elements.
Despite training on pairs of screenshots and code, DL-based approaches are still restricted by their training data, which may contain limited web elements and programming knowledge. In a nutshell, existing studies are unsuitable for building real-world software websites.

Multimodal Large Language Models (MLLMs) have seen significant advancements recently~\cite{Dai2023InstructBLIPTG, openai_gpt4, google_gemini_api}, offering an alternative solution for design-to-code. By integrating image processing capabilities into large language models (LLMs), MLLMs have demonstrated superior ability than CNNs for understanding images and answering visual questions~\cite{Yang2023TheDO}. Furthermore, research has shown that LLMs have remarkable performance on various code intelligence tasks~\cite{Hou2023LargeLM}, including code generation~\cite{Yu2023CoderEvalAB, Li2023EnablingPT, Du2023ClassEvalAM, Dong2023SelfcollaborationCG, Jiang2023SelfEvolveAC, Gilbert2023SemanticCW}, code completion~\cite{Nijkamp2022CodeGenAO, Ding2023ASE, Dibia2022AligningOM, Li2023NuancesAT, Li2022CCTESTTA, Chen2021EvaluatingLL}, and code summarization~\cite{Mastropaolo2021StudyingTU, Mastropaolo2022UsingDL,Gu2022AssembleFM, Chen2022OnTT, Gao2023ConstructingEI, Arakelyan2023ExploringDS}. 
The strong image understanding and code generation capabilities of MLLMs showcase their potential to transform website \add{design images} into functional web code.
% Recent work has demonstrated the potential of directly using MLLMs to generate UI code from screenshots~\cite{Si2024Design2CodeHF}.

Unfortunately, notwithstanding MLLMs' impressive performance in image-to-text tasks, directly applying them for UI code generation results in unsatisfactory performance because of the complicated nature of GUI. Unlike generating general descriptions, translating images into UI code poses unique challenges for MLLMs. Firstly, it must accurately detect and classify diverse elements and nested structures within a single webpage, such as buttons, text fields, and images. Secondly, it demands the precise replication of intricate layouts and styles. Modern GUIs cannot be constructed merely through the hardcoded positioning of texts, images, and controls; rather, the model must possess a thorough understanding of a GUI framework’s components, including the supported visual effects and compositions, to reproduce elements with the details (e.g., colors, fonts, margins) and positioning~\cite{Chen2018FromUI}. Thirdly, unlike static descriptions, the generated code must be executable, adhering to the syntax and semantic requirements of front-end frameworks.

To understand MLLMs's capability in transforming UI images to UI code, we conduct a motivating study to identify three main issues in this process: (1) element omission, where certain visual components are missing in the generated code, (2) element distortion, where elements are inaccurately reproduced in terms of shape, size, or color, and (3) element misarrangement, meaning the elements are not positioned or ordered correctly relative to their design layout. In contrast, when we cropped the inaccurately generated images into smaller pieces and asked MLLMs to re-generate UI code for each segment, we observed enhanced performance on these smaller, cropped UI images.
\textit{In short, the smaller and more focused the image, the better the resulting code quality.}

Building on this insight, we propose a novel \textbf{\underline{D}}ivide-and-\textbf{\underline{C}}onquer-based method to \textbf{\underline{Gen}}erate website code from UI designs, namely \textbf{\methodname}. This end-to-end approach follows the principles of a typical divide-and-conquer algorithm: breaking down a complex problem into smaller parts, solving each part individually, and then combining the solutions to address the original problem.
Specifically, \methodname consists of two phases: \textit{division} and \textit{assembly}. The first phase involves slicing the \add{design images} into smaller, more manageable, yet semantically meaningful pieces, and generating the HTML and CSS code for each distinctive segment. The division aligns with the real-world front-end development practices~\cite{csschopper2024} and is achieved via a novel image segmentation algorithm. 
\methodname then generates code for each segment, followed by recursively reassembling the code to reconstruct the entire website. This reassembly process is the reverse of the division procedure, where code from smaller, child segments is progressively integrated to build up their parent segments. This assembly continues until the full website structure is restored.

Our study curated a dataset comprising 348 top-ranked real-world websites. We evaluate the effectiveness of our methodology across several cutting-edge MLLMs, using 111 webpages sampled from the dataset. \methodname achieved up to 8\% improvement in visual similarity \add{and 15\% improvement in code similarity for large input images }compared to other design-to-code methods. We further demonstrate that \methodname is robust \add{against complex webpages}. Moreover, \methodname is shown to generalize well across different MLLMs, demonstrating the effectiveness of the divide-and-conquer methodology. Human evaluations show that DCGen can produce more visually similar target webpages than competing methods and can help developers implement webpages faster.

% effectively boost the performance of MLLMs, leading to a maximum of 14\% performance gain on visual similarity of the original website and generated website compared to other direct promoting methods and is robust to variations in website complexity, being able to maintain superior performance across various website complexities. Furthermore, experiments also show that \methodname can be generalized to different MLLMs, thereby affirming the effectiveness of our approach. 

In summary, our paper makes the following contributions:
\begin{itemize}
    \item We initiated a motivating study to uncover types of errors in the MLLM-powered design-to-code process, revealing the importance of visual segments in generation quality.
    \item We propose DCGen with a divide-and-conquer approach. It starts with generating code for individual image segments, and then merges these solutions to produce the complete UI code.
    \item The experiments on real-world webpages demonstrate the superiority of DCGen over other methods in both visual similarity and code similarity. Human evaluations also confirm that DCGen can help developers implement webpages faster and more similar to the UI designs.
    \item We implement DCGen into a user-friendly demo tool and release all datasets and code implementation for future research in this field.
\end{itemize}

\section{Background}
% In this section, we introduce concepts related to the front-end development process and the problem definition of design-to-code throughout the paper.

\paragraph{Concepts in Front-End Development}
% \subsubsection{Front-end Development}
Front-end development concentrates on creating the user interface and enhancing the user experience of websites. It involves Hypertext Markup Language (HTML) for content structuring, Cascading Style Sheets (CSS) for styling, and optionally JavaScript for interactivity and dynamic webpage features.
% \subsubsection{HTML Elements \& Tags}
HTML is a markup language used in front-end web development to structure webpages. Tags in HTML define the document's structure, like \texttt{<h1>} to \texttt{<h6>} for different heading levels, and \texttt{<p>} for paragraphs. Each HTML element includes an opening tag, content, and a closing tag, forming the basic block of a webpage. 
% Below is a simplified HTML snippet containing six different elements:
% \begin{lstlisting}[frame=single, basicstyle=\ttfamily\footnotesize, xleftmargin=1em, xrightmargin=1em,]
% <html>
% <head>
%     <title>Sample Page</title>
% </head>
% <body>
%     <h1>Welcome to My Website</h1>
%     <p>This is a paragraph of text.</p>
% </body>
% </html>
% \end{lstlisting}
% \subsubsection{CSS}
CSS is used in front-end web development to style webpages and enhance the appearance of HTML elements. It allows developers to define unique styles for elements through selectors, properties, and values. 
% For example, to change the color and font size of all \texttt{<h1>} elements, one developer might write:

% \begin{lstlisting}[frame=single, basicstyle=\ttfamily\footnotesize, xleftmargin=1em, xrightmargin=1em,]
% h1 {
%     color: blue;
%     font-size: 24px;
% }
% \end{lstlisting}

\paragraph{Problem Definition}
This work addresses the design-to-code task, where the input is a visual design of a webpage, and the goal is to generate HTML+CSS code that accurately reproduces it. Let $I_0$ be the design image of a webpage, and $C_0$ be the ideal HTML+CSS code. Given $I_0$, an MLLM $M$ generates code $C_g = M(I_0)$, and the rendered output $I_g$ of $C_g$ should closely match $I_0$ in both structure and appearance.

Because design files are typically unavailable, we use screenshots of existing webpages as proxies for design images $I_0$, and use the corresponding HTML+CSS code as ground truth $C_0$. This allows us to construct datasets for evaluating the model’s ability to recover code from visual designs. The quality of $C_g$ is assessed based on (1) code similarity to $C_0$—matching elements and structure—and (2) visual similarity between the rendered output $I_g$ and the design image $I_0$.

% Let $C_0$ be a file of HTML+CSS code of a webpage, $I_0$ be the screenshot of the webpage, and $M$ be an MLLM. The \textit{design-to-code} task takes the image $I_0$ as input and outputs a file of generated HTML+CSS code $C_g = M(I_0)$ that approximate $C_0$. Figure~\ref{fig:framework} illustrates an example of the input and output. The quality of $C_g$ is evaluated based on both functional similarity and visual similarity. Specifically, $C_g$ should be functionally similar to $C_0$, meaning it consists of a similar set of HTML elements and a comparable nested structure of the elements. Additionally, $I_g$, the screenshot of the webpage rendered from $C_g$, should be visually similar to $I_0$.

\section{Dataset Collection}
% \subsection{Dataset Construction}
\label{subsec:dataset-construction}
This section illustrates how we develop a dataset containing representative websites for this study.
We randomly sample websites from the top 500 listed on the Tranco\footnote{\url{https://tranco-list.eu/}. Accessed in Jan 2024 }~\cite{Pochat2018TrancoAR}, and process them. First, we filter out all invalid websites (e.g., blocked, empty, or requiring human verification). 
Next, to make the websites self-contained and eliminate external dependencies that affect the UI code generation, we save each one into a single HTML file using the SingleFile toolkit\footnote{https://github.com/gildas-lormeau/SingleFile}, remove all external links, and replace images with placeholders. Finally, we clean the websites by removing all elements that do not impact the website's appearance (i.e., hidden elements).
The dataset collection yields 348 websites containing an average of 26 unique tags, where each HTML code file has an average length of 99,431 tokens.

Table~\ref{tab:dataset_stat} provides quantitative metrics to evaluate the complexity and comprehensiveness of our dataset. \textbf{(1) Length:} We tokenize the HTML code files in each category using the GPT-2 tokenizer. The average number of tokens is 99,431, which suggests that our dataset covers a wide range of HTML code complexity. \textbf{(2) Total number of tags:} We counted the number of tags in HTML files, finding an average of 961, which provides insights into the structural complexity of the documents. \textbf{(3) DOM tree depth:} We calculated the Document Object Model (DOM) tree depth in HTML files. The average depth is 17, indicating a high nesting complexity of HTML tags. \textbf{(4) Number of unique tags:} We also counted the number of unique tags, with an average of 26, reflecting the diversity and content richness of the HTML files.

% Then, we process the data following the same procedures as in \ref{subsubsec:pilot-data-sample}. 

\begin{table}[t]
    \centering
    \begin{minipage}{0.47\textwidth}
        \footnotesize
        \centering
        \caption{Statistics of the dataset.}
        % \vspace{-0.1in}
        \label{tab:dataset_stat}
        \begin{tabular}{lrrr}
        \toprule
         & \textbf{Min} & \textbf{Max} & \textbf{Average} \\
        \midrule
        Length (tokens) & 383 & 868,108 & 99,431 \\
        Tag Count       & 2 & 16,331 & 961 \\
        DOM Depth       & 2 & 40 & 17 \\
        Unique Tags     & 2 & 66 & 26 \\
        \midrule
        \textbf{Total size}      &   &    &{348} \\
        \bottomrule
        \end{tabular}
    \end{minipage}%
    \hfill
    \begin{minipage}{0.47\textwidth}
        \footnotesize
        \centering
        \caption{Distribution of common mistakes (one element can have multiple mistakes).}
        \vspace{-0.1in}
        \label{tab:pilot_data}
        \begin{tabular}{lrr}
            \toprule
            \textbf{Category} & \textbf{Percentage (\%)} & \textbf{Count} \\
            \midrule
            Omission       & 85.34 & 1,450 \\ % missing, lack of content
            Distortion     & 2.56  & 44   \\ % wrong color, wrong scale
            Misarrangement & 12.71 & 216  \\ % wrong layout, wrong position
            Correct        & 2.35  & 40   \\
            \midrule
            \textbf{Total} &       & 1,699\\
            \bottomrule
        \end{tabular}
    \end{minipage}
    \vspace{-0.1in}
\end{table}

\section{Motivating study}
\label{sec:motivating-study}
We begin with a motivating study to examine MLLMs' ability to generate UI code from screenshots. Taking GPT-4o as an example, we showcase three common mistakes MLLMs make during the generation and further discuss how to mitigate these errors.

\subsection{Study Preparation}
To explore MLLM's performance in the design-to-code task, we compare the original website's appearance with the version generated by MLLM (i.e., GPT-4o in this section).
This involves an image-based comparison to check if the \textit{visual elements} from the original are accurately reproduced in the MLLM-generated website, after rendering. 

\subsubsection{Dataset sampling.} 
The pilot study samples 12 top-ranked websites from the dataset described in Section~\ref{subsec:dataset-construction}, such as Amazon, Apple, and Facebook. After sampling, we prompt GPT-4o to generate website code from the screenshot using the direct prompt described in Section~\ref{subsubsec:prompt}.

\subsubsection{Visual Element Locating (for original websites).}
This step identifies the bounding boxes for each element (tag) in the code within the webpage screenshot, referred to as visual elements.
Using Selenium WebDriver\footnote{https://selenium-python.readthedocs.io/}, we extract the bounding box coordinates for all elements and eliminate any bounding boxes that intersect.
This collection of bounding boxes enables us to locate all HTML elements in a website's screenshot.
Overall, these webpages contain a total of 1,699 HTML elements, with individual webpages featuring between 4 and 434 elements, averaging 142 elements per site.

% \begin{figure}
%     \centering
%     \includegraphics[width=0.5\columnwidth]{Sections/figs/annotation.png}
%     \caption{Annotation tool. Annotators are provided with an original webpage (left) with a bounding box specifying the visual element and a generated webpage (right). 
%     }
%     \label{fig:annotation-tool}
%     \vspace{-15pt}
% \end{figure}

\subsection{What failures does MLLM produce when converting design to code?}
We recruit two graduate students in our university as annotators to assess the accuracy of visual elements in MLLM-generated webpages compared to the original ones.  The annotators are provided with pairs of screenshots, with the original on the left and the MLLM-generated version on the right. A red bounding box highlights the visual element in the original screenshot.
Annotators are asked with two tasks: 
(1) Draw the bounding box around the corresponding element in the MLLM-generated screenshot. (2) Describe any differences between the generated visual element and the original in short phrases. 

After two annotators independently complete these tasks, we take the union of the two annotators' descriptions as the final annotation for each element and manually analyze the descriptions.

\begin{figure}[t]
    \centering
    \begin{minipage}{0.28\textwidth}
        \centering
        \begin{subfigure}[b]{0.462\columnwidth}
            \centering
            \includegraphics[width=\textwidth]{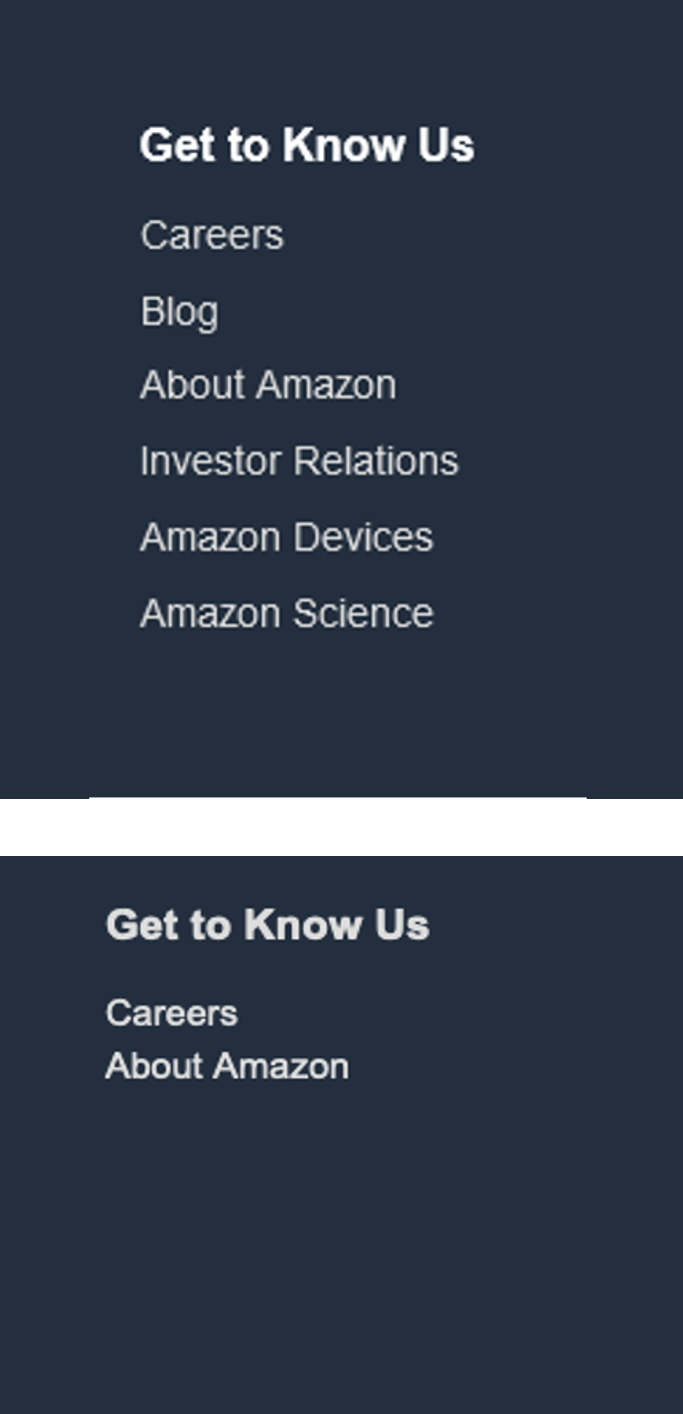}
            \caption{Omission}
            \label{fig:motivating-omission}
        \end{subfigure}
        % \hfill
        \begin{subfigure}[b]{0.478\columnwidth}
            \centering
            \includegraphics[width=\textwidth]{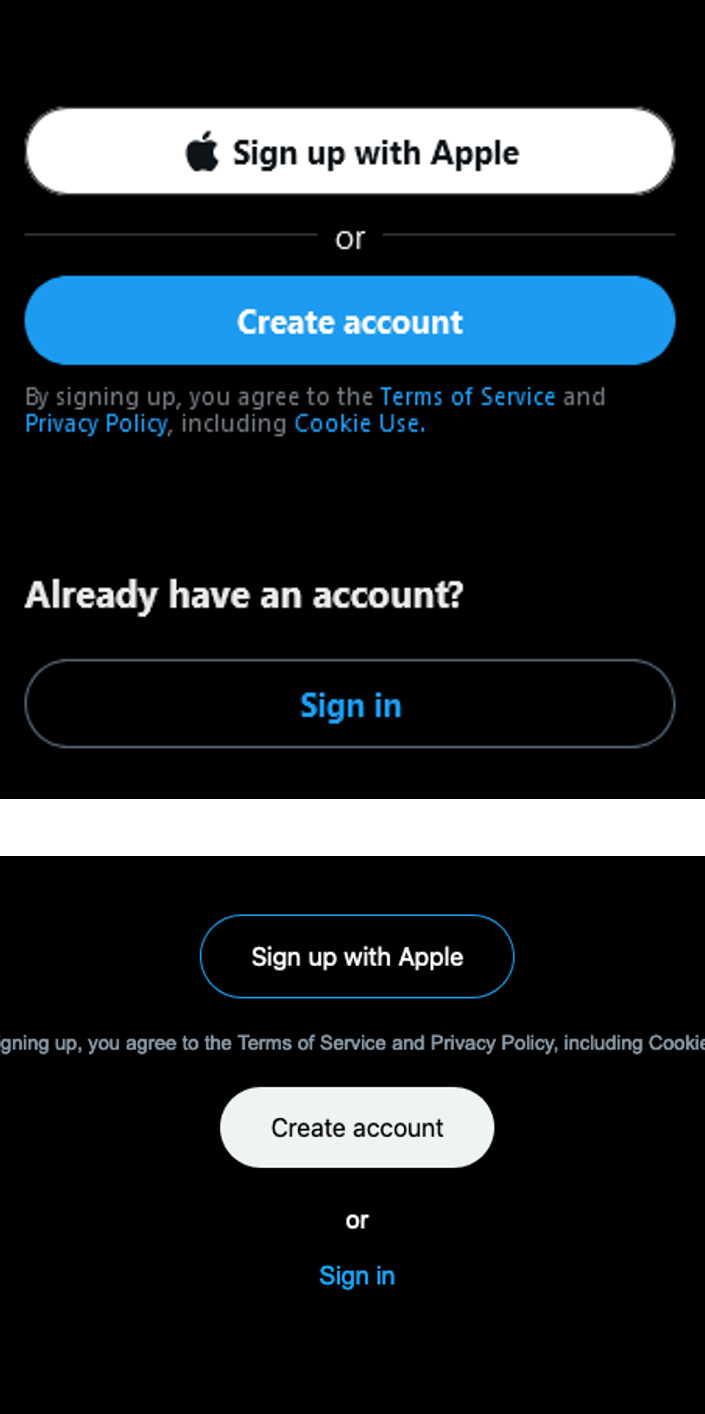}
            \caption{Distortion}
            \label{fig:motivating-distortion}
        \end{subfigure}
        \hfill
      
        \begin{subfigure}[b]{0.82\columnwidth}
            \centering
            \includegraphics[width=\textwidth]{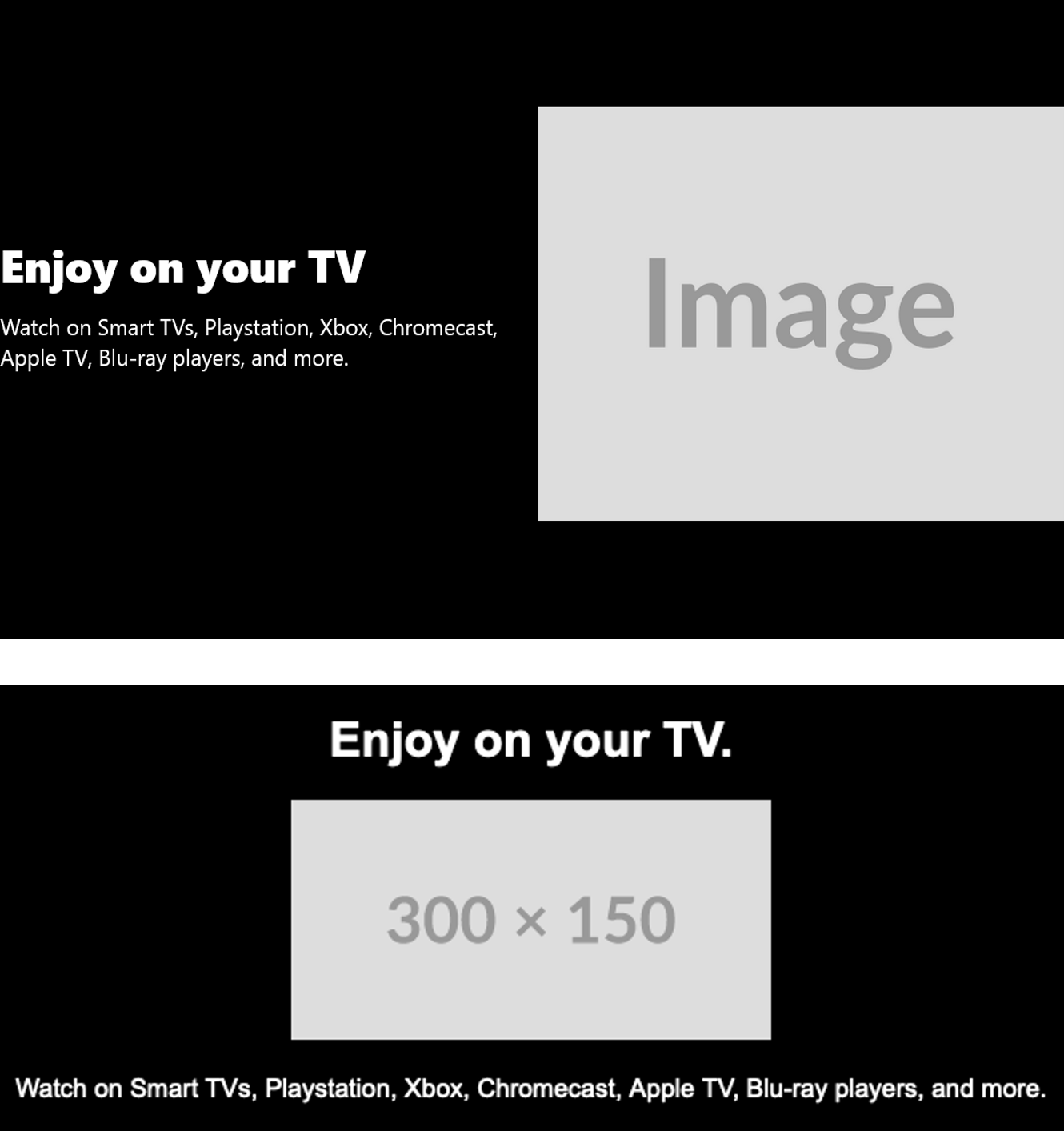}
            \caption{Misarrangement}
            \label{fig:motivating-misarrangement}
        \end{subfigure}
        \caption{Examples of error cases (bottom of each subfigure).}
        \label{fig:pilot-example-1}
    \end{minipage}
    % \hfill
    \begin{minipage}{0.68\textwidth}
        \centering
        \sbox{\arrangebox}{%
          \begin{subfigure}[b]{0.48\columnwidth}
          \centering
          \includegraphics[width=\textwidth]{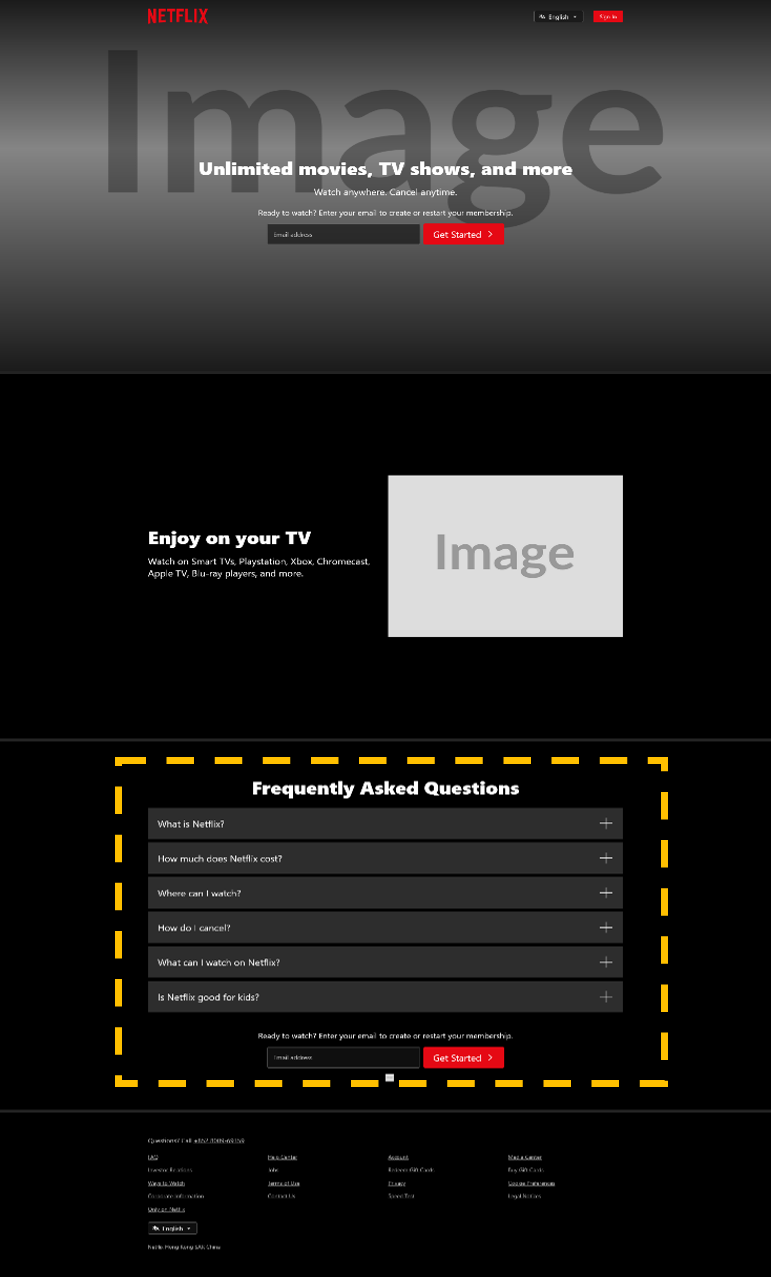}
          \caption{Input. The full image is the input of (b) and the yellow dashed line area is the input of (c).}
          \end{subfigure}%
        }
        \setlength{\arrangeht}{\ht\arrangebox}
        \usebox{\arrangebox}\hspace{3px}
        \begin{minipage}[b][\arrangeht][s]{0.48\columnwidth}
          \begin{subfigure}[t]{\textwidth}
          \centering
          \includegraphics[width=\textwidth]{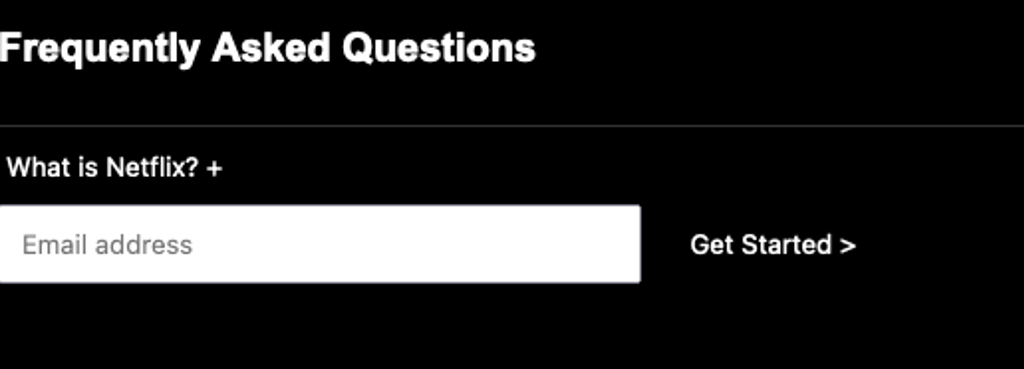}
          \caption{Generated yellow dashed line area when input is the full screenshot.}
          \end{subfigure}\vspace{5px}
          \begin{subfigure}[b]{\textwidth}
          \centering
          \includegraphics[width=\textwidth]{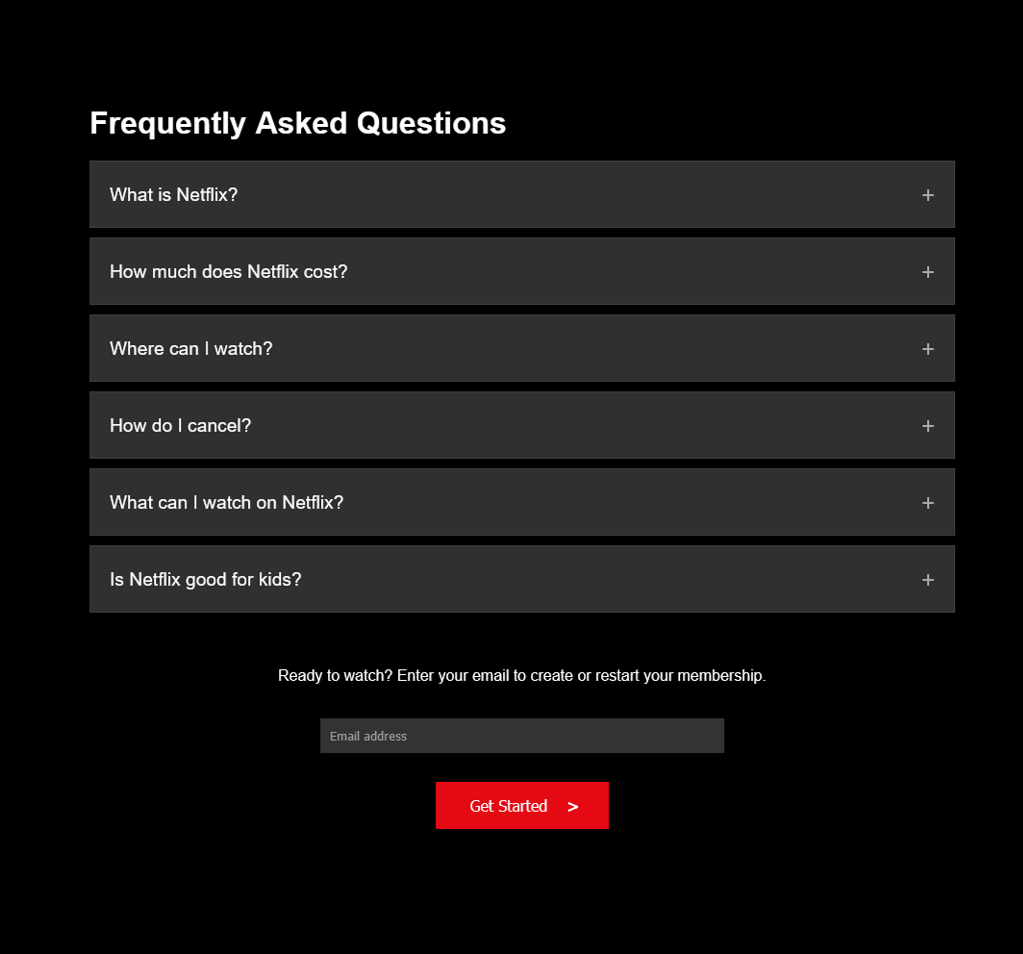}
          \caption{Generated yellow dashed line area when input is the area's screenshot.}
          \end{subfigure}
        \end{minipage}
        % \vspace{-8pt}
        \caption{An example of the experiment in the motivating study.}
        \label{fig:pilot-example}
    \end{minipage}
    \vspace{-15pt}
\end{figure}

\textit{Result Analysis.}
After inspection, we discover that all the differences between the generated element and the original element can be characterized into three categories, that is, element omission, element distortion, and element misarrangement, and one element can have multiple errors at the same time. 
Figure~\ref{fig:pilot-example-1} provides an example of each type of error, with the original website displayed above and the MLLM-generated version below in each subfigure.

\begin{enumerate}[leftmargin=*]
    \item Element omission: Certain visual elements are missing in the generated code. As shown in Figure~\ref{fig:motivating-omission}, the text elements below ``About Amazon'' are missing.
    \item Element distortion: Elements are reproduced inaccurately in terms of shape, size, or color. Figure~\ref{fig:motivating-distortion} illustrates a case where the ``sign up with Apple'' button is incorrectly colored.
    \item Element misarrangement: Elements are not positioned or ordered correctly relative to their adjacent elements or the overall design layout. In Figure~\ref{fig:motivating-misarrangement}, MLLM misplaces the text ``Enjoy on your TV'' in relation to the image.
\end{enumerate}

The distribution of these errors is shown in Table~\ref{tab:pilot_data}, showing that element omission accounts for the majority of error cases, which is potentially due to the visual shortcomings of MLLMs~\cite{Tong2024EyesWS}.
In total, MLLM correctly reproduces only 40 out of 1,699 elements, highlighting the need for a practical UI code generation approach.

\subsection{Can we correct mistakes by visual segments?}
After noting the unsatisfactory performance in generating precise UI code, we investigate the reasons for these errors. LLMs are known to excel at simple, small-step tasks, and struggle with long-term planning for complex tasks~\cite{wei2022chain}.
To address this, existing studies have developed the Chain-of-Thought paradigm or LLM Agents, which decompose complex assignments into smaller, manageable sub-tasks, enabling models to perform more reasoning steps ~\cite{feng2024prompting}.

Inspired by these studies, we break down full screenshots into smaller image segments to investigate whether MLLMs perform better with simpler inputs. This approach helps us determine whether the model is fundamentally incapable of generating the content or if its mistakes can be alleviated.
In specific, we conduct follow-up investigations on two failure cases.

The first case focuses on omitted elements. We isolate each missing element by cropping the screenshot to include only that specific element and then request the model to generate the corresponding code. 
Figure~\ref{fig:pilot-example} depicts this process, where Figure~\ref{fig:pilot-example}b is the Frequently Asked Questions (FAQ) area generated from the full screenshot, and Figure~\ref{fig:pilot-example}c is the FAQ area generated from the cropped screenshot. The results show that with these isolated inputs, the model successfully generates the previously omitted content.  This demonstrates that the model is intrinsically capable of producing the required outputs but struggles with complex inputs.

The second investigation concerns distorted and misarranged elements. We compare the image similarity between the original elements and the model-generated elements from both full and cropped screenshots. We measure image similarity using the CLIP embeddings~\cite{Radford2021LearningTV}, as outlined in~\cite{Si2024Design2CodeHF}. The findings reveal an overall improvement in similarity scores from 73.7\% to 76.0\% by providing the cropped screenshot, confirming that smaller and more focused images can enhance the quality of the generated code.

\begin{tcolorbox}[colback=gray!20, colframe=gray!20, width=\columnwidth]
\textbf{Insights:} 
While existing MLLMs struggle with generating accurate UI code, we observe that breaking down full screenshots into smaller visual segments improves performance. This decomposition allows models to conduct more reasoning steps, each focused on a manageable sub-generation task.
\end{tcolorbox}

\begin{figure*}[t]
    \centering
    \includegraphics[width=\textwidth]{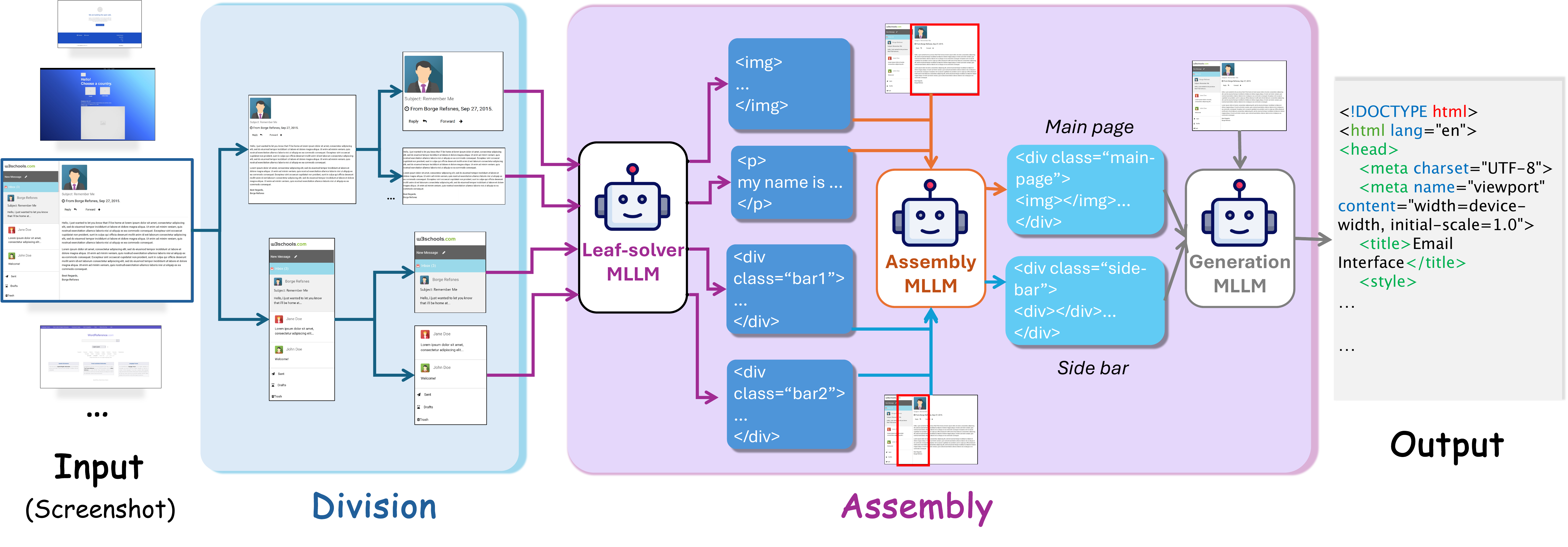}
    % \vspace{-25pt}
    \caption{The framework of DCGen.}
    \label{fig:framework}
    % \vspace{-15pt}
\end{figure*}

\section{Methodology}
In this section, we present \methodname, an end-to-end framework for automated UI code generation from design images. Motivated by the traditional ``\textit{divide and conquer}'' algorithmic technique, \methodname decomposes a complicated screenshot into smaller, doable visual segments, solves each part individually, and then combines the solutions for the original problem.

\methodname accepts website screenshots and outputs a file of UI Code. Figure~\ref{fig:framework} shows the overall workflow of \methodname, consisting of two stages: \textit{division} and \textit{assembly}. 
In the division stage, \methodname searches for horizontal and vertical separation lines in an interleaved fashion and splits the image accordingly, breaking the image into smaller pieces until no more separation lines are found or a user-defined maximum depth is reached. At this point, \methodname passes each image segment to MLLMs for code generation. Afterward, in the assembly stage, code from smaller, child segments is progressively integrated to build up their parent segments. This recursive assembly continues until the full structure of the website is restored.

% This process is carried out recursively, dividing the image into smaller yet complete pieces until no more separation lines can be found in the segments or a user-defined maximum depth is reached. At this stage, \methodname will pass each image segment to MLLMs for code generation.

% \begin{figure}[t]

% \end{figure}

\subsection{Division}
\label{subsec:division}
Given a full screenshot, the division stage outputs a hierarchical structure of image separations, with each leaf node representing a fully subdivisioned image (i.e., the smallest segment). The screenshot division faces two challenges. First, the division algorithm must carefully examine visual elements to ensure that no single element is split into multiple segments. Second, the image segments should be stored in order, so that they can be reconstructed for the full webpage later.

The following section delineates a novel division algorithm, which comprises two main components: separation line detection and screenshot subdivision. The separation line detector first identifies all explicit or implicit separation lines within screenshots (either vertical or horizontal).
Then, the screenshot subdivision component leverages these lines to segment the image into smaller and hierarchical segments.
The two components work alternately, dividing the image horizontally, then vertically for the subimages.
% Figure~\ref{fig:subdivisions} illustrates various outcomes of the screenshot segmentation algorithm at different depths (i.e., the number of alternations).
% To resolve the first challenge, we propose a separation line detector that identifies all explicit or implicit separation lines within screenshots (either vertical or horizontal).
% Then, the image subdivision component leverages these lines to segment the image into smaller and hierarchical segments, tackling the second challenge. 

\subsubsection{Separation Line Detection}

\methodname starts with identifying all separation lines (either vertical or horizontal) in a full screenshot. These separation lines must divide the screenshot into rectangular and meaningful pieces, where each of the pieces contains elements that are as complete as possible; otherwise, it would be more difficult to combine code for fractured pieces later. Traditional image segmentation techniques concentrate on distinguishing irregularly shaped objects from the background~\cite{minaee2021image}, becoming unsuitable for dividing webpages.

\begin{figure}[t]

    \begin{minipage}{.6\linewidth}
    \scriptsize
    \begin{algorithm}[H]
    \caption{Separation Line Detection Algorithm}
    \label{algo:separation-line}
    \begin{algorithmic}[1]
        \REQUIRE img, var\_thr, diff\_thr, portion\_thr, window\_size
        \STATE lines $\leftarrow$ []
        \FOR{$i \leftarrow window\_size + 1$ \TO $\text{len}(img) - 1$}
            \STATE upper $\leftarrow$ img[i - window\_size - 1]
            \STATE window $\leftarrow$ img[i - window\_size : i]
            \STATE lower $\leftarrow$ img[i]
            \STATE var $\leftarrow$ variance(window)
            \STATE is\_blank $\leftarrow$ (var < var\_thr)
            \STATE diff\_top $\leftarrow$ abs\_diff(upper, window[0])
            \STATE is\_border\_top $\leftarrow$ portion(diff\_top > diff\_thr)  > portion\_thr
            \STATE diff\_bottom $\leftarrow$ mean\_abs\_diff(lower, window[-1])
            \STATE is\_border\_bottom $\leftarrow$ portion(diff\_bottom > diff\_thr) > portion\_thr
            \IF{is\_blank \AND (is\_border\_top \OR is\_border\_bottom)}
                \IF{is\_border\_bottom}
                    \STATE pos $\leftarrow$ i 
                \ELSE
                    \STATE pos $\leftarrow$ i - window\_size 
                \ENDIF
                \STATE lines.append(pos)
            \ENDIF
        \ENDFOR
        \STATE \RETURN sorted(lines)
    \end{algorithmic}
    \end{algorithm}
\end{minipage}
\begin{minipage}{0.35\textwidth}
\centering
    \begin{subfigure}[b]{0.6\textwidth}
        \centering
        \includegraphics[width=\textwidth]{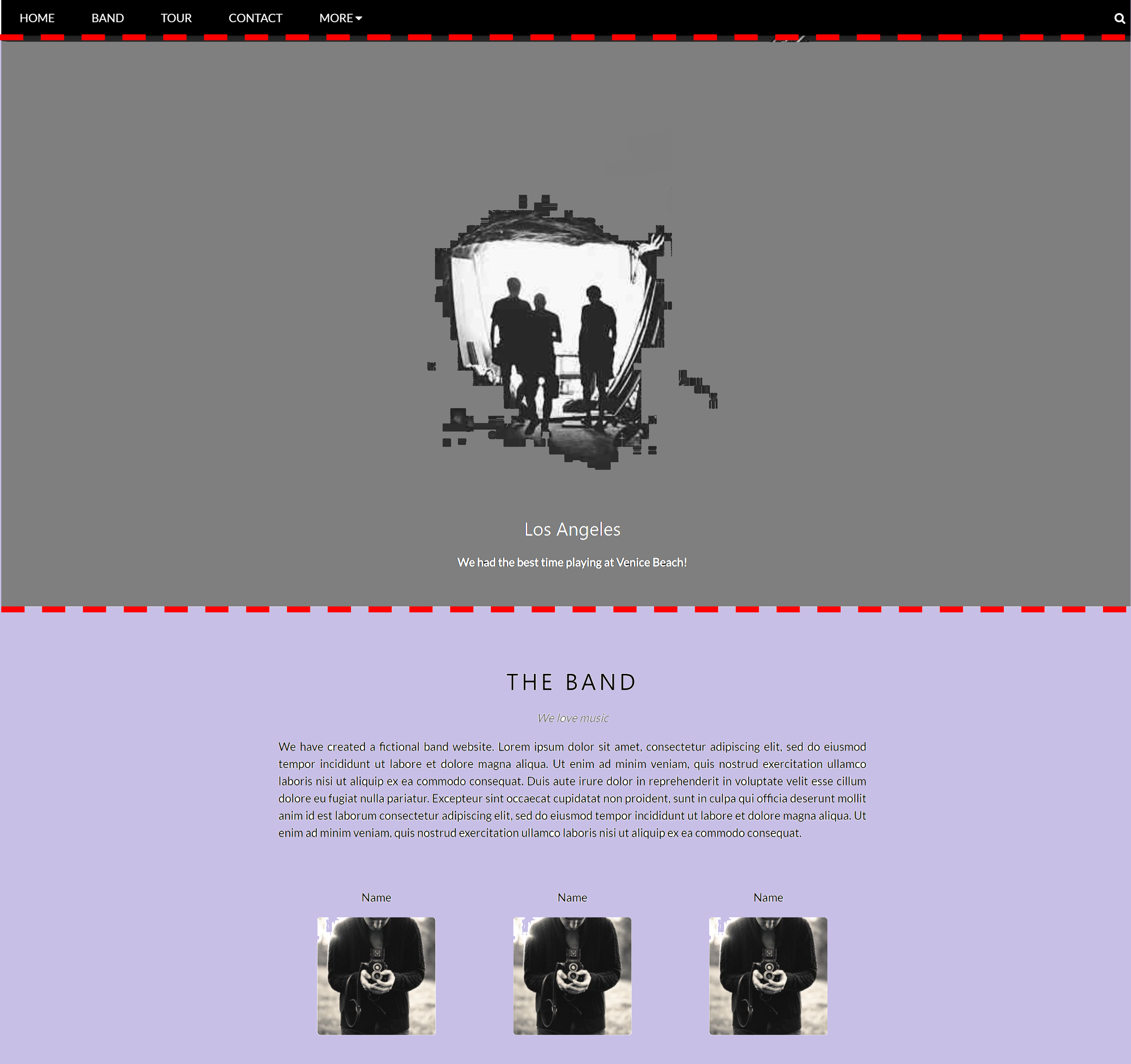}
        \caption{Explicit.}
        \label{fig:sub1}
    \end{subfigure}
 
    \begin{subfigure}[b]{0.6\textwidth}
        \centering
        \includegraphics[width=\textwidth]{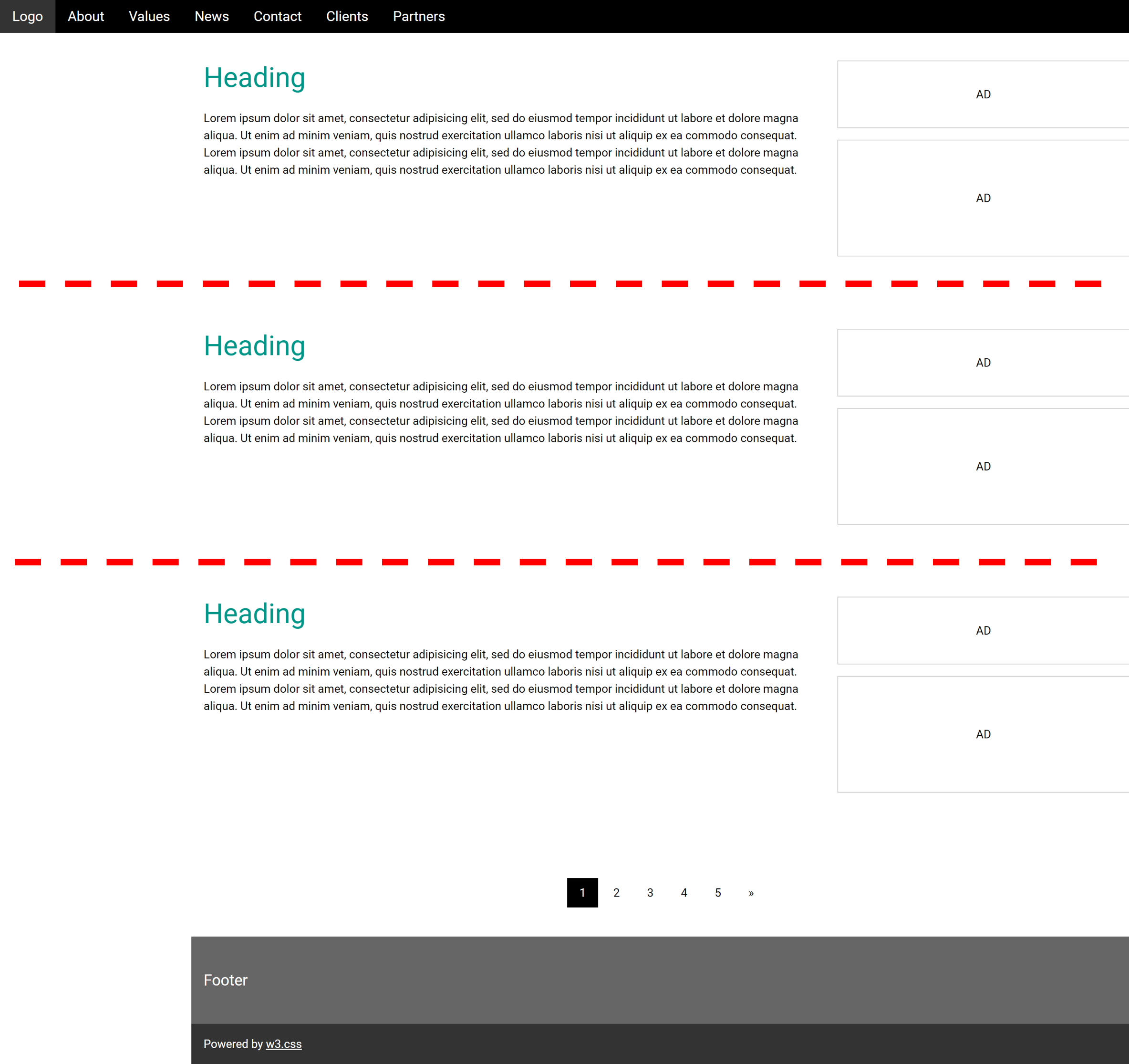}
        \caption{Implicit.}
        \label{fig:sub3}
    \end{subfigure}

    \caption{Separation lines (red dot lines).}
    \label{fig:separation-line}
\end{minipage}

\end{figure}

We observe that websites contain explicit (visible lines separating content) or implicit (blank spaces or borders between content elements) separation lines to organize different sections and contents.
Figure~\ref{fig:separation-line} depicts examples of both types. 
Figure~\ref{fig:separation-line}a employs explicit lines to separate the grey area from the purple area, while Figure~\ref{fig:separation-line}b adapts implicit lines to arrange text sections.

Motivated by two types of separation lines, our algorithm 
 (shown in Algorithm~\ref{algo:separation-line}) begins by converting the screenshot to grayscale, and then scans the screenshot row by row to identify explicit lines and implicit lines, respectively. The separation lines are signified by sudden shifts in pixel values.

% It \remove{then} scans the image row by row to identify sudden shifts in pixel values\replace{ that may}, which signify the presence of separation lines. The algorithm is illustrated in Algorithm~\ref{algo:separation-line}.

For detecting implicit separation lines, our approach incorporates a sliding window technique (lines 3-5), analyzing groups of consecutive rows to detect zones of minimal brightness variation, indicating potential blank spaces (lines 6-7). The algorithm compares each window against the rows directly above and below (lines 8-11). A separation line is identified if (1) the variance within the window is lower than a set threshold, suggesting a blank area, and (2) the brightness difference between this window and the rows directly above and below exceeds another threshold over more than \textit{portion\_thr} of the row's length (line 9, 11). This dual-threshold method facilitates the identification of subtle borders and blanks, recognizing both the low-content areas and the significant transitions at their edges. The selection of thresholds and parameters are detailed in Section 7.1. 

For detecting explicit separation lines, the method is equivalent to finding implicit separation lines by setting the sliding window size to one. In particular, the algorithm scrutinizes each row for uniformity in pixel brightness. A row with variability below a predefined threshold is flagged as a potential line. Confirmation of a separation line occurs if the average difference in brightness between this row and its predecessor or successor surpasses a set threshold over at least \textit{portion\_thr} of the row.

% \begin{figure}[t]
% \begin{minipage}{0.56\textwidth}
%     \centering
%     \begin{subfigure}[b]{0.32\columnwidth}
%         \centering
%         \includegraphics[width=\textwidth]{Sections/figs/example.png}
%         \caption{Original image}
%         \label{fig:div-0}
%     \end{subfigure}
%     \hfill
%         \begin{subfigure}[b]{0.32\columnwidth}
%         \centering
%         \includegraphics[width=\textwidth]{Sections/figs/example_cut1.png}
%         \caption{Depth = 1}
%         \label{fig:div-1}
%     \end{subfigure}
%         \hfill
%     \begin{subfigure}[b]{0.32\columnwidth}
%         \centering
%         \includegraphics[width=\textwidth]{Sections/figs/example_cut2.png}
%         \caption{Depth = 2}
%         \label{fig:div-2}
%     \end{subfigure}
%     \caption{Examples of image subdivisions of different depths.}
%     \label{fig:subdivisions}
% \end{minipage}

% \vspace{-5pt}
% \end{figure}

\subsubsection{Screenshot Subdivision}
 We divide the screenshot into smaller segments by \textit{alternately and recursively} applying horizontal and vertical divisions to mirror real-world development practices~\cite{csschopper2024}. 
 This process is applied recursively to each resulting segment until no further separation lines can be identified or until a user-defined maximum recursion depth is reached. 
 
 % Figure~\ref{fig:subdivision} illustrates this subdivision procedure.

We employ a tree structure to store the hierarchical screenshot divisions. If the image $I$ is divided into segments $I_a$ and $I_b$, then $I$ is considered the parent of segments $I_a$ and $I_b$, with $I_a$ and $I_b$ as its children. Consequently, the number of recursions corresponds to the number of layers in the tree. Each leaf node represents a fully subdivisioned screenshot, which is the smallest segment.

% \begin{figure}
%     \includegraphics[width=0.6\textwidth]{Sections/figs/subdivision.png}
%     \vspace{-10pt}
%     \caption{Screenshot subdivision example.}
%     \label{fig:subdivision}
%     \vspace{-15pt}
% \end{figure}

\subsection{Assembly}
\label{subsec:assembly}
In the assembly process, we integrate the solutions of image subdivisions to generate complete UI code for the original screenshot. The assembly process is essentially the reverse of the division procedure, where code from smaller child segments is progressively integrated to build up their parent segments until the full structure of the website is restored. 

\add{We propose two different implementations of the assembly process, agent-based and rule-based. Agent-based assembly is shown in Figure~\ref{fig:framework}. \methodname first provides the MLLM (i.e., Leaf-solver MLLM) with leaf image segments, then asks it to generate the code of the segments. Then, \methodname provides the MLLM (i.e., Assembly MLLM) a parent segment and all the code generated from its children's image segments. In the last step (merging for root node), \methodname calls the Generation MLLM to produce the complete UI code for full screen based on the code from its children segments and the entire screenshot. We adapt the basic instructions based on the previous study by Si et al.~\cite{Si2024Design2CodeHF}. All the prompt details for the leaf-solver MLLM, assembly MLLM, and generation MLLM are made available in our artifact. Algorithm~\ref{algo:dcgen} shows the pseudocode of the complete DCGen algorithm.}

\add{For rule-based assembly, instead of using multiple MLLMs to assemble partial code output at different depths, we replace the assembly step with a rule-based program that uses CSS grid for layout positioning\footnote{\url{https://www.w3schools.com/css/css_grid.asp}}. As usual, we first divide the input image into multiple segments. The position of each segment within the overall layout is recorded. Then, MLLMs are used only at the leaf level to independently generate HTML and CSS code for each segment. Consequently, instead of using LLMs to assemble the segments into a full webpage, we directly place each generated segment into a CSS grid layout at the recorded positions. The assembled template, now containing all the fine-grained code blocks in the correct positions, is passed to a generator LLM for final adjustments, ensuring cohesion, flexibility, and accurate visual details. Algorithm~\ref{algo:dcgen_efficient} shows the pseudocode of this implementation.}

\begin{algorithm}[t]
\scriptsize
\caption{\add{DCGen-Agent}}
\label{algo:dcgen}
\begin{algorithmic}[1]
\STATE \textbf{Input:} $img$, $bot$, $depth$, $max\_depth$, $leaf\_prompt$, $node\_prompt$
\IF{$depth \geq max\_depth$}
    \RETURN $bot.generate\_code(leaf\_prompt, img)$
\ENDIF
\STATE $img\_cuts \gets segment\_image(img)$
\IF{no cuts found}
    \RETURN $bot.generate\_code(leaf\_prompt, img)$
\ENDIF
\STATE $code\_parts \gets []$
\FOR{each $cut$ in $img\_cuts$}
    \STATE $code\_parts \gets$ run in new thread: $dcgen(cut, bot, depth + 1, max\_depth, leaf\_prompt, node\_prompt)$
\ENDFOR
    \STATE \RETURN $bot.generate\_code(node\_prompt + code\_parts, img)$
\end{algorithmic}
\end{algorithm}

\begin{figure}

    \begin{minipage}{0.5\textwidth}
        \begin{algorithm}[H]
        \scriptsize
        \caption{\add{DCGen-Rule}}
        \label{algo:dcgen_efficient}
        \begin{algorithmic}[1]
        \STATE \textbf{Input:} $img$, $bot$
        \STATE $img\_segs,\space seg\_positions \gets segment\_image(img)$
        \STATE $code\_segs \gets generate\_code\_parallel(img\_segs,\space bot)$
        \STATE $HTML\_template \gets CSS\_grid(seg\_positions, codes\_segs)$
        \STATE $final\_code \gets generate\_code(HTML\_template, img, bot)$
        \STATE \RETURN $final\_code$
        \end{algorithmic}
        \end{algorithm}
    \end{minipage}
    \begin{minipage}{0.46\textwidth}
    \vspace{-10pt}
        \footnotesize
        \centering
        \caption{Statistics of the test data}
        \label{tab:testset_stat}
        \begin{tabular}{lrrr}
        \toprule
         & \textbf{Min} & \textbf{Max} & \textbf{Average} \\
        \midrule
        Length (tokens) & 383 & 123,578 & 28,430 \\
        Tag Count       & 2 & 396 & 188 \\
        DOM Depth       & 2 & 28 & 12 \\
        Unique Tags     & 2 & 57 & 19 \\
        \midrule
        \textbf{Total size}      & & & 111 \\
        \bottomrule
        \end{tabular}
        
    \end{minipage}
\end{figure}

\section{Experiment}
% \subsection{Research Questions}
We evaluate the effectiveness of \methodname for the design-to-code performance by answering the following research questions.
\begin{itemize}[leftmargin=*]
    \item \add{\textbf{RQ1:} How do different parameters influence the performance of the separation algorithm?}
    \item \textbf{RQ2:} How does \methodname compare to various methods in design-to-code?
    \item \textbf{RQ3:} How robust is \methodname for complex designs?
    \item \textbf{RQ4:} How generalizable is \methodname for other state-of-the-art (SOTA) MLLMs?
    \item \textbf{RQ5:} What is human's feedback on \methodname in real-world application development workflows?
\end{itemize}

\subsection{Experiment Setup}
\subsubsection{Backbone Models} To assess the performance of \methodname, we employ it to evaluate four SOTA MLLMs: GPT-4o~\cite{openai_gpt4o}, Gemini-1.5-flash and \add{Geimini-2-flash}~\cite{google_gemini_api}, Claude-3-Sonnet~\cite{anthropic_claude}. Access to all models is facilitated through their respective official APIs. The specific model numbers are 20240513 for GPT-4o, 20240229 for Claude-3-Sonnet, Gemini-1.5-flash-latest accessed during May 2024, and Gemini-2.0-flash accessed Feb 2025. Due to a limitation on computational resources and budget, we randomly sampled 111 websites from our dataset to form the testing set. Table~\ref{tab:testset_stat} shows the statistics of the testing dataset. The data is rich in HTML tags and diverse in overall complexity. 

\subsubsection{Parameter Setting} For MLLM model configurations, we set the temperature to 0 and the $max\_tokens$ parameter to 8192 for each model, while other parameters are maintained at their default settings as specified in the corresponding API documentation~\cite{google_gemini_api_docs, openai_vision_guide, anthropic_vision_docs}. For parameters in the image segmentation algorithm detailed in Section 5.1, \add{we set \textit{var\_thr=50}, \textit{diff\_thresh=45}, \textit{diff\_portion=0.9}, \textit{window\_size=50} and set the division depth to 2}.

\subsubsection{Efficiency}
 \add{DCGen leverages parallelism to promote efficiency, where all queries at the same depth are processed simultaneously, as shown in Algorithm~\ref{algo:dcgen}. Given an MLLM response time of $t$ per query, the total response time for DCGen is $(depth + 1)*t = 3t$ for $depth=2$. As for cost efficiency, on each depth where MLLM produces partial code for an image segmentation, the sum of the output code on each depth is roughly upper bound by the length of the final generated code $L$. Thus, the total token usage (and cost) for agent-based assembly is upper bounded by $(depth + 1)*L = 3L$ for $depth = 2$. In short, DCGen costs three times that of direct querying. However, with the rule-based assembly, DCGen is able to eliminate intermediate MLLM queries by generating code at the leaf and root levels. In such a way, the optimal cost can be lowered to $2L$, which stays the same as self-refinement prompting.}

% \begin{figure}[t]
%     \centering
%     \includegraphics[width=\columnwidth]{Sections/figs/setup_prompt.pdf}
%     \caption{Propmt for baseline approaches.}
%     \label{fig:prompting-exp}
% \end{figure}

\subsubsection{Baselines} \label{subsubsec:prompt} \add{We first compare our method with the Pix2Code method~\cite{Beltramelli2018Pix2c} that trained deep-learning model to convert screenshot to code. Then we compare various prompting strategies including }Chain-of-Thought (CoT) prompting~\cite{Wei2022ChainOT} that generates a chain of thought for each question and then generates the corresponding code. For CoT, we use the "let’s think step by step" instruction from Chae et. al~\cite{Chae2024LanguageMA}. Self-refine prompting~\cite{Chen2023TeachingLL} let the model refine its own generated code via multi-turn conversation. We adopt the self-refine prompting and direct promoting method from Si et al.~\cite{Si2024Design2CodeHF}. The details of prompting strategies used in our experiment as baselines are all available in our artifact.

% We do not select previous deep-learning-based approaches as baselines since they are either tailored to the mobile app environment~\cite{Chen2018FromUI, Nguyen2015ReverseEM, Moran2018MachineLP}, not opensourced~\cite{Xu2021Image2e}, do not take website screenshot as input or do not produce complete website code~\cite{Asiroglu2019AutomaticHC, Beltramelli2018Pix2c}.
\subsection{Automatic Metrics}
Our experiments use screenshots of existing webpages as proxies for design images $I_0$, and use their corresponding HTML+CSS code as ground truth $C_0$. We evaluate the performance of \methodname in both high-level similarity and fine-grained element matching. 

\subsubsection{High-Level Similarity} 
% We break high-level similarity down to both code similarity and image similarity.
We utilize two high-level similarities in this paper including \textbf{code similarity} and \textbf{image similarity}, respectively.

\add{For code similarity, we adopt the Normalized Levenshtein Distance (edit distance)~\cite{rapidfuzz_levenshtein} to evaluate the closeness of generated code $C_g$ to the original code $C_0$. Levenshtein Distance 
is a robust method that is widely used to measure similarity between two sequences~\cite{Haldar2011LevenshteinDT}. The score is calculated as S = $1 - distance/(l_1+l_2)$~\cite{rapidfuzz_levenshtein}, which provides a direct measurement of similarity between the generated and ground truth HTML, with a higher value indicating higher similarity.}

For image similarity, we measure image similarity between $I_0$ and $I_g$ via CLIP Score ~\cite{Si2024Design2CodeHF}. CLIP score relies on the similarity of their CLIP \cite{Radford2021LearningTV} embeddings, denoted as CLIP($I_0$,  $I_g$). Specifically, we extract features using CLIP-ViT-B/32.

\subsubsection{Fine-Grained Measurements}
\label{sec:fine-grained}
High-level metrics capture only the overall similarity of images and code, lacking the detail necessary to analyze specific model performances. To address this limitation, we adopt a suite of element-matching metrics for evaluating generated webpages in terms of text content, position, and color introduced by Si et al. \cite{Si2024Design2CodeHF}. Given reference and generated webpage screenshots \( I_0 \) and \( I_g \), the algorithm detects visual element blocks and matches them using the Jonker-Volgenant algorithm. It then evaluates the similarity of the matching blocks across several aspects: (1) Block-match measures the ratio of matched block sizes to all block sizes, penalizing missing or hallucinated blocks; (2) Text similarity calculates character overlap using the Sørensen-Dice coefficient; (3) Color similarity employs the CIEDE2000 formula to evaluate perceptual color differences; and (4) Position similarity assesses the alignment of block centers.

\section{Results \& Analysis}

\subsection{RQ1. How do parameters influence the performance of the separation algorithm?}

\add{In this section, we examine how five hyper-parameters, i.e., \textit{diff\_thresh}, \textit{var\_thresh}, \textit{diff\_portion}, \textit{window\_size}, and \textit{depth}, affect the performance of the separation algorithm. To evaluate the precision of the separation algorithm, we extract bounding boxes formed by the separation lines and use the standard Intersection over Union (IoU) metric~\cite{Redmon2015YouOL} to quantifies the overlap between predicted and ground-truth bounding boxes. The IoU between two bounding boxes $b_1,b_2$ is calculated as $IoU=\frac{\text{Intersection}(b_1,b_2)}{\text{Union}(b1,b2)}$, with higher value indicating more overlapping or higher precision.}

\add{For each parameter configuration, we assess the algorithm's performance using our entire test set. Specifically, given an input image, we recursively apply our separation algorithm to generate separation lines. The rectangular areas defined by these lines serve as predicted bounding boxes. We then extract ground-truth bounding boxes from the corresponding HTML code using JavaScript. For each predicted bounding box, we compute its maximum IoU with all ground-truth bounding boxes. The average IoU across the test set provides a measure of the algorithm’s precision. However, it is important to note that IoU serves only as a reference metric. The goal of the separation algorithm is not to perfectly match each HTML element but to generate reasonable divisions that help MLLMs focus on finer details.}

\add{Beyond IoU, we also evaluate the extent of image division. Some parameter settings may yield highly accurate but overly coarse segmentations (e.g., producing only a few bounding boxes), while others may result in more granular yet less precise divisions. To quantify this, we introduce the $\text{Separation Rate} = 1 - \frac{b_{max}}{S}$, where \( b_{max} \) represents the size of the largest undivided leaf bounding box, and \( S \) is the total image size. A higher separation rate indicates finer image segmentation.}

\subsubsection{Effect of \textit{diff\_thresh}, \textit{var\_thresh}, and \textit{diff\_portion}.}
\add{To analyze the impact of \textit{diff\_thresh}, \textit{var\_thresh}, and \textit{diff\_portion}, we fixed \textit{depth}=2 and \textit{window\_size}=50, then selected three \textit{diff\_portion} values (0.3, 0.6, and 0.9)  spanning its range (0,1). For each \textit{diff\_portion}, we plotted the IoU scores across different \textit{diff\_thresh} and \textit{var\_thresh} settings.}

\begin{figure}[t]
\vspace{-0.1in}
  \begin{subfigure}{0.72\textwidth}
    \centering
    \includegraphics[width=\linewidth]{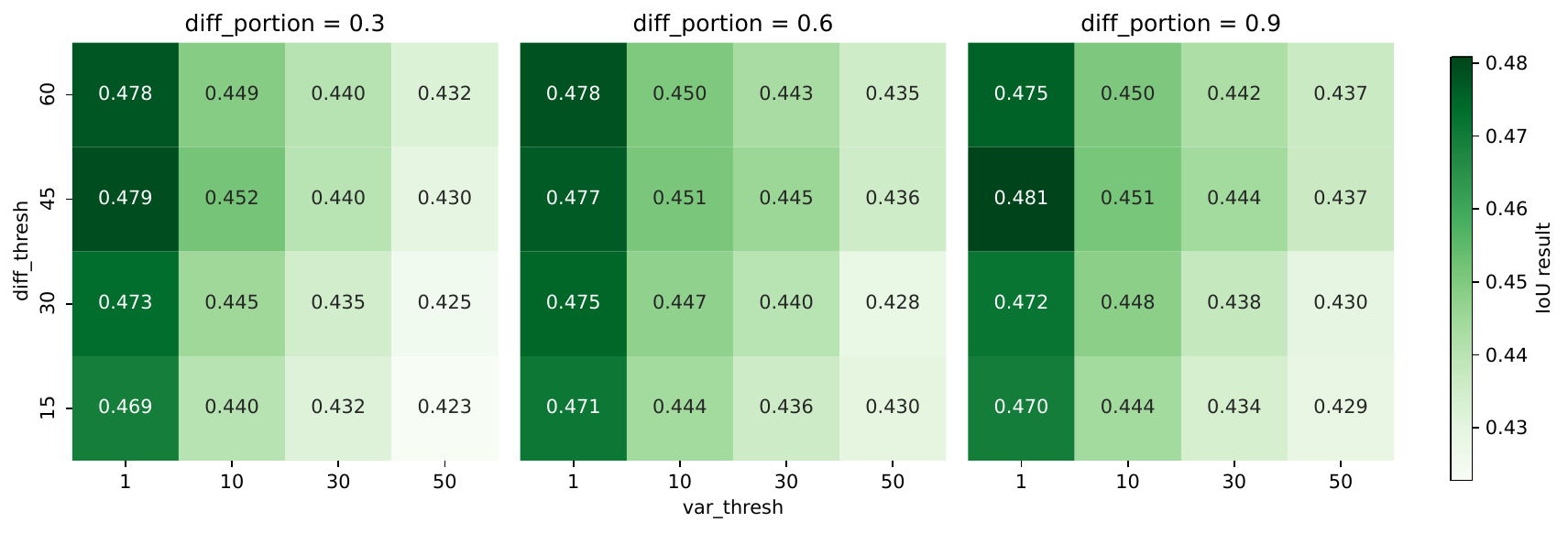}
    \caption{}
    \label{fig:diff-portion}
  \end{subfigure}%
  \begin{subfigure}{0.27\textwidth}
    \centering
    \includegraphics[width=\linewidth]{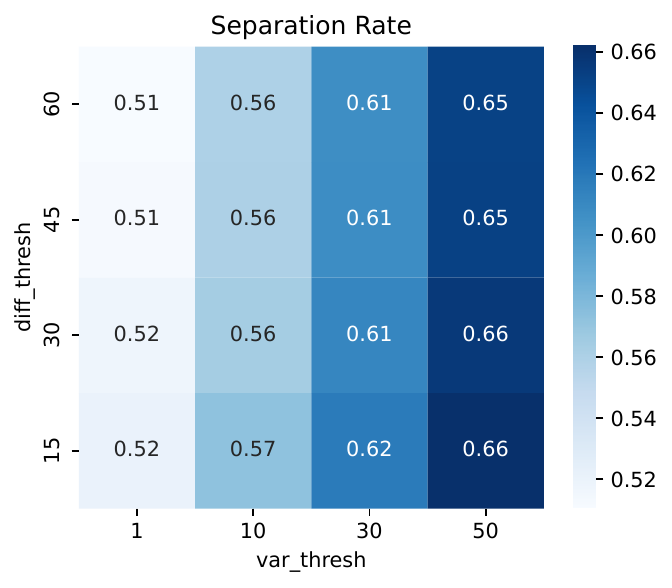}
    \caption{}
    \label{fig:diff-portion-sep}
  \end{subfigure}
  \vspace{-0.1in}
  \caption{\add{(a) Effect of \textit{diff\_portion}, \textit{diff\_thresh}, and \textit{var\_thresh} on the separation algorithm's IoU score. (b) Effect of \textit{diff\_thresh}, \textit{var\_thresh} on separation rate when \textit{diff\_portion}=0.9. Larger \textit{var\_thresh} improves separation rate.}}
  \label{fig:thresh}
\end{figure}

\add{Figure~\ref{fig:diff-portion} shows that \textit{diff\_portion} and \textit{diff\_thresh} have minimal impact ($\le1\%$) on algorithm precision, whereas a lower \textit{var\_thresh} improves accuracy. Experiments indicate that IoU variations across parameter settings remain within approximately 5\%, demonstrating the separation algorithm's robustness. These fluctuations arise because lower \textit{var\_thresh} reduces the likelihood of misclassifying areas as ``blank'', thereby limiting unnecessary separation proposals, which results in a more conservative approach, yielding fewer but more accurate separation lines. Meanwhile, as illustrated in Figure~\ref{fig:diff-portion-sep}, increasing \textit{var\_thresh} significantly enhances the separation rate, producing finer-grained segmentations.}

\subsubsection{Influence of Window Size and Depth.}
\begin{figure}[t]
  \begin{subfigure}{0.72\textwidth}
    \centering
    \includegraphics[width=\linewidth]{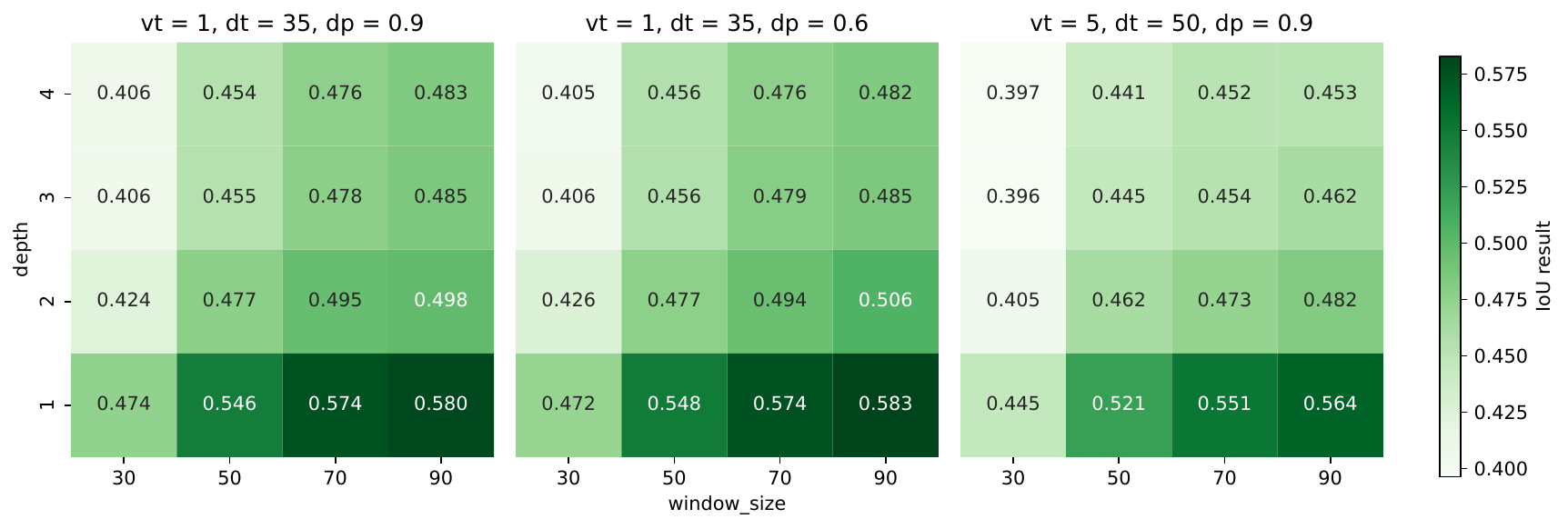}
    \caption{}
    \label{fig:window-depth}
  \end{subfigure}%
  \begin{subfigure}{0.27\textwidth}
    \centering
    \includegraphics[width=\linewidth]{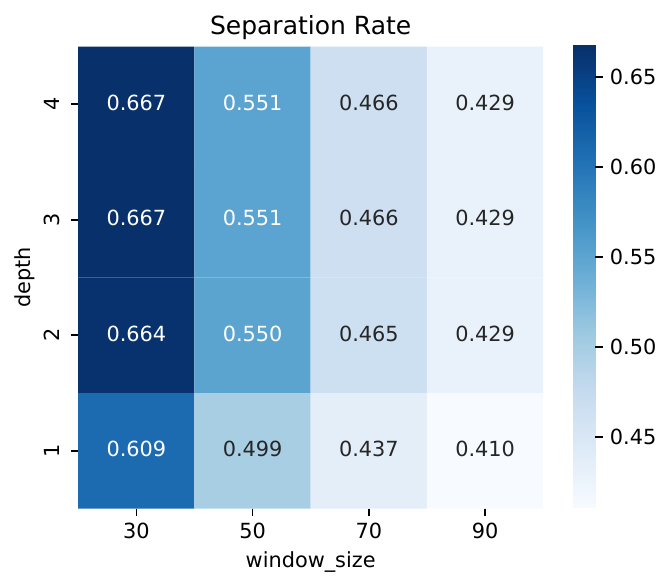}
    \caption{}
    \label{fig:window-depth-sep}
  \end{subfigure}
  \vspace{-0.1in}
  \caption{\add{(a) IoU under different window sizes and depths across three sets of \textit{var\_thresh} (vt), \textit{diff\_thresh} (dt), and \textit{diff\_portion} (dp). (b) Separation rate under different window sizes and depths when vt=5, dt=50, and dp=0.9. A larger depth and smaller window size lead to a finer separation rate. Depth 2, 3, and 4 have a similar separation ratio.}}
  \label{fig:window}
\end{figure}

\add{To analyze the effect of sliding window size and separation depth on the separation algorithm, we selected three sets of parameters from Section 7.1.1 and plotted the average IoU across different window sizes and depths. The results, shown in Figure~\ref{fig:window-depth}, indicate that a smaller depth and a larger window size generally lead to more accurate separation.}

\add{On the other hand, we observe a similar trade-off between precision and separation rate. Figure~\ref{fig:window-depth-sep} shows that a larger depth and smaller window size lead to a finer separation rate. However, while an increase from 1 to 2 enhances the separation rate, further increases in depth barely influence the rate.}

\subsubsection{Parameter Selection} 
\begin{figure}[htbp]
    \centering
    \begin{subfigure}{0.29\textwidth}
        \centering
        \includegraphics[width=\linewidth]{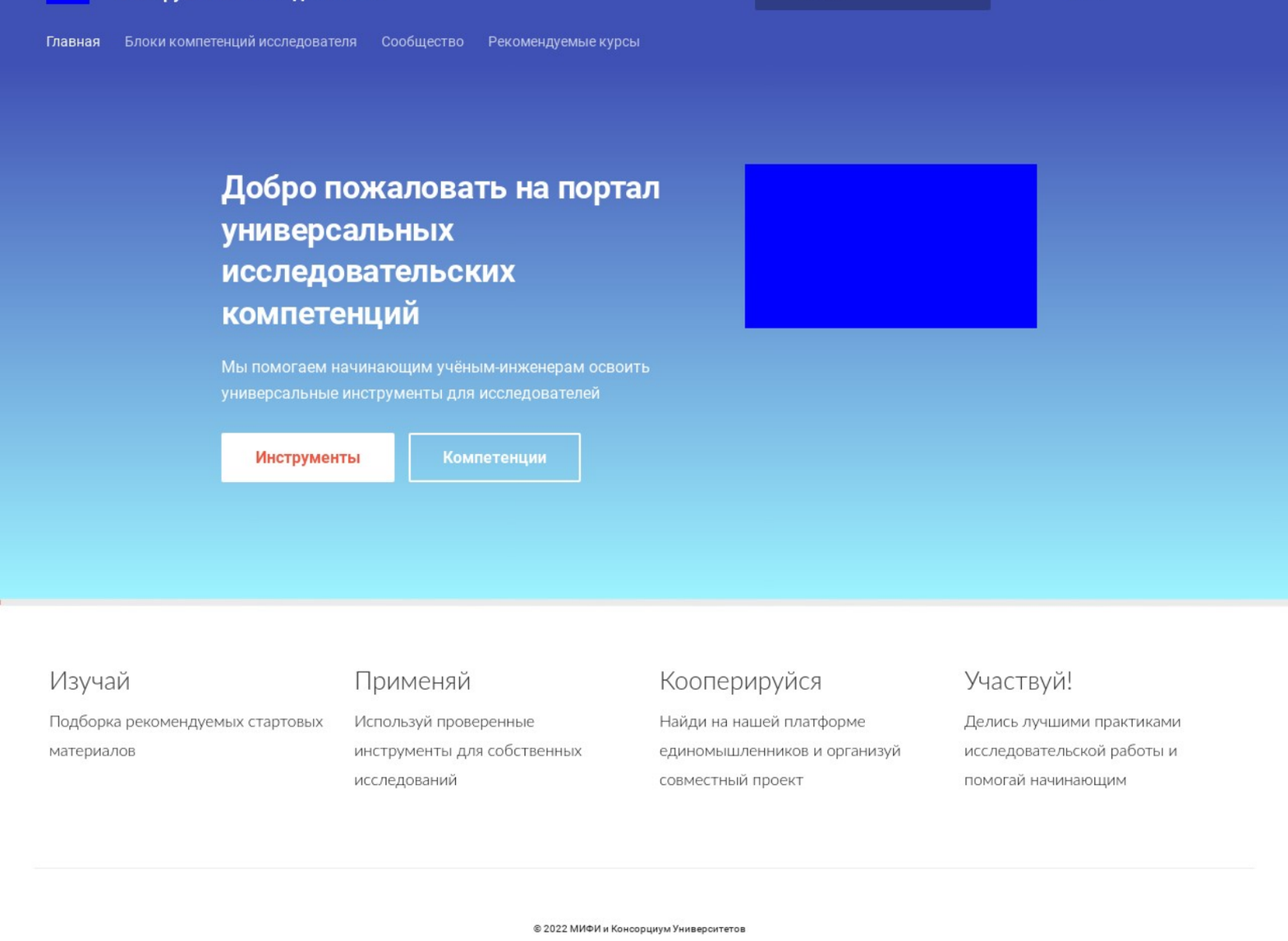}
        \caption{Input image}
        \label{fig:compare-1}
    \end{subfigure}
    \hfill
    \begin{subfigure}{0.27\textwidth}
        \centering
        \includegraphics[width=\linewidth]{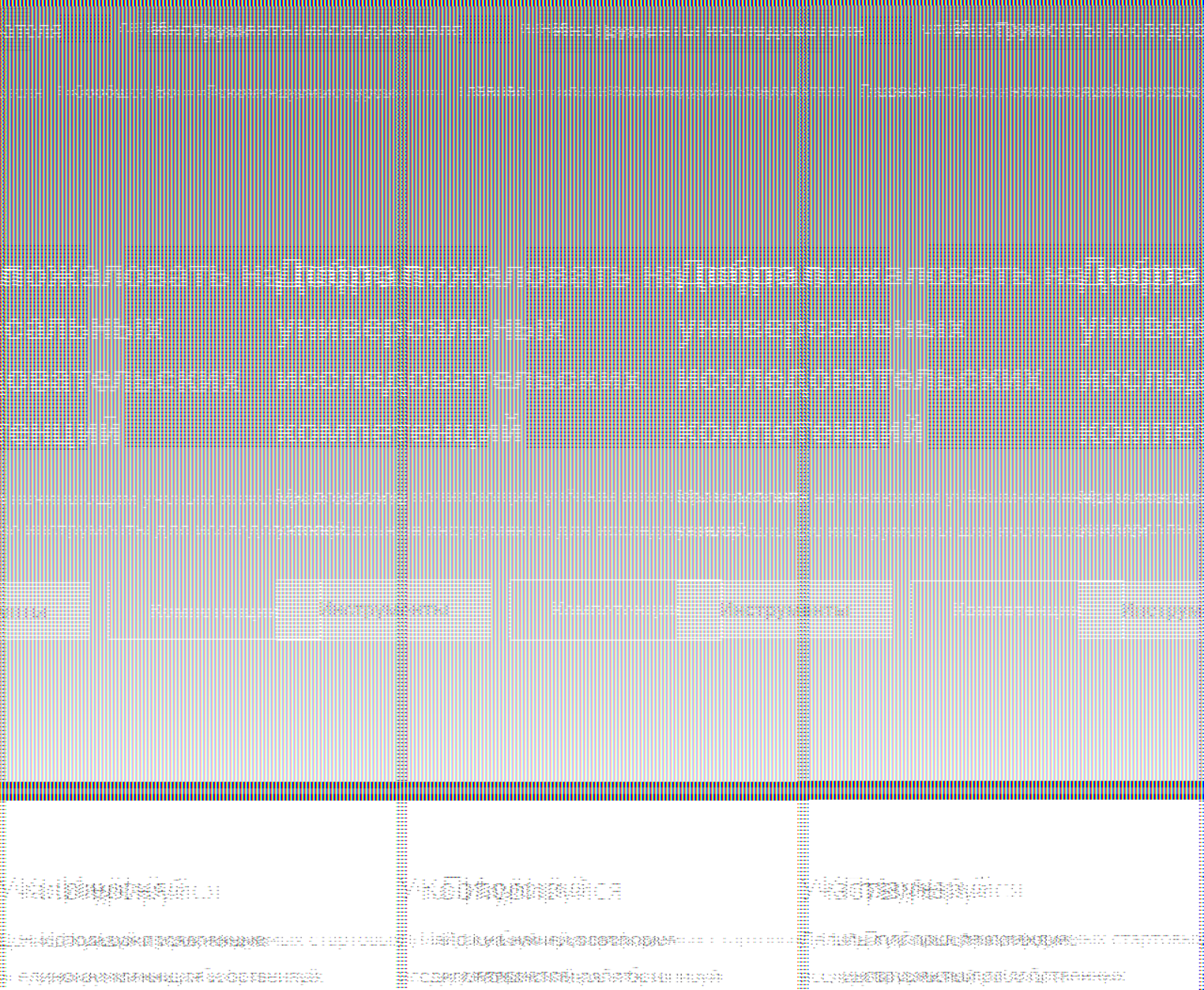}
        \caption{IoU=0.58, Sep Rate=0.45}
        \label{fig:compare-2}
    \end{subfigure}
    \hfill
    \begin{subfigure}{0.27\textwidth}
        \centering
        \includegraphics[width=\linewidth]{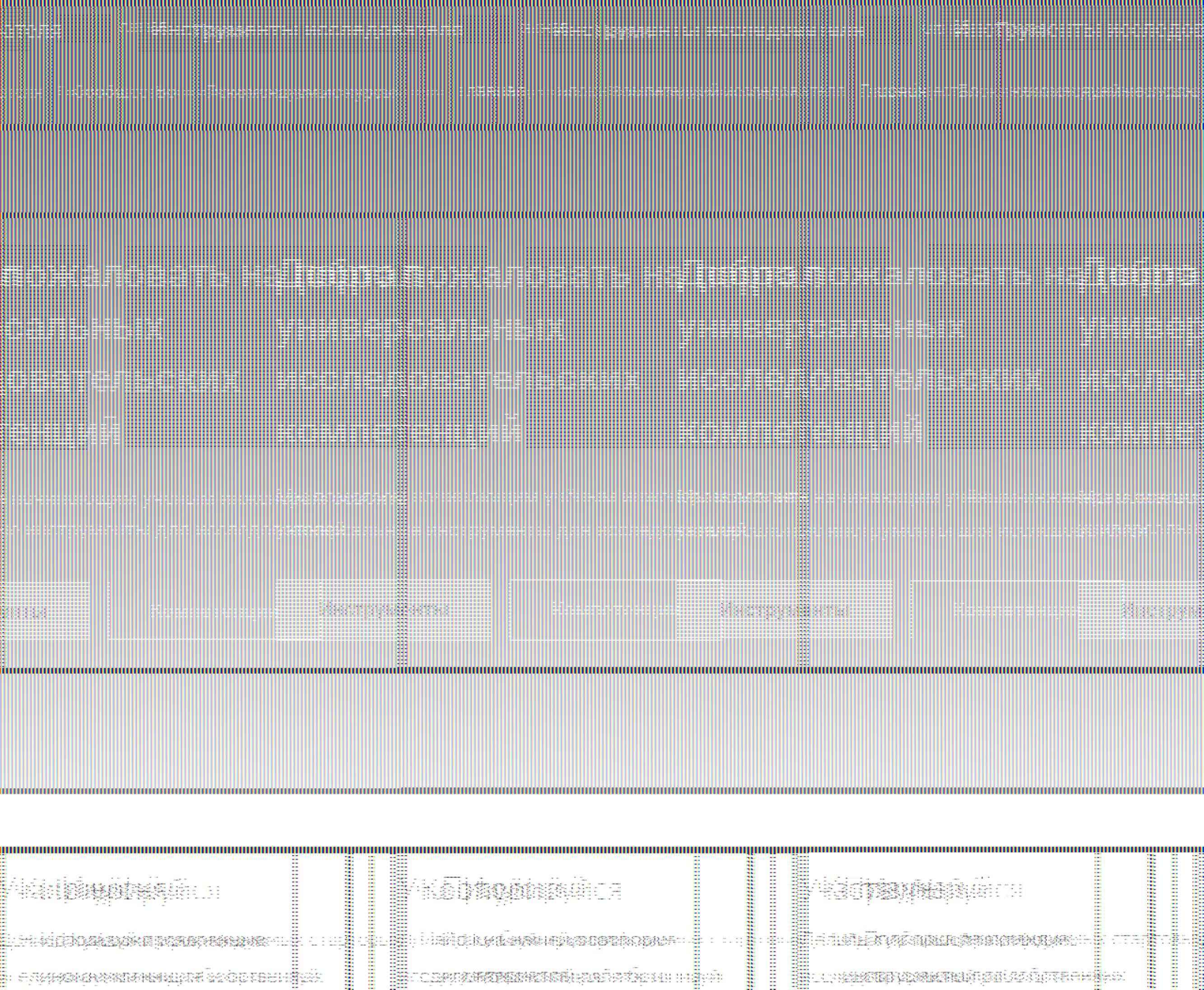}
        \caption{IoU=0.43, Sep rate=0.61}
        \label{fig:compare-3}
    \end{subfigure}
    \caption{\add{Comparison of different parameter settings for the separation algorithm. (b) ws=90, vt=1, dt=45, dp=0.9, depth=1. A larger \textit{ws}, smaller \textit{vt} and \textit{depth} result in fewer but more accurate segmentation areas (c) ws=50, vt=50, dt=45, dp=0.9, depth=2. A smaller \textit{ws}, larger \textit{vt} and \textit{depth} lead to finer but less precise segmentation (these are the parameters we used in the final experiment).}}
    \label{fig:compare}
\end{figure}

\add{We selected the parameters based on observations from Sections 7.1.1 and 7.1.2. Observing that \textit{diff\_portion} and \textit{diff\_thresh} have minimal influence on algorithm precision, we randomly pick \textit{diff\_portion}=0.9 and \textit{diff\_thresh}=45 for implementing DCGen. While smaller \textit{var\_thresh} improves separation accuracy, it fluctuates by only about 5\%. On the other hand, larger \textit{var\_thresh} significantly improves the separation rate. Therefore, we pick a larger \textit{var\_thresh}=50. Then, to achieve a better separation rate and maintain a reasonable IoU, we pick \textit{window size}=50. Finally, since an increase of depth from 1 to 2 obviously enhances the separation rate, further increases in depth barely influence the rate but decrease IoU significantly, we pick depth=2 to achieve a balance between precision and division rate. Figure~\ref{fig:compare} shows the trade-off we observed in Sections 7.1.1 and 7.1.2, and Figure~\ref{fig:compare-3} shows an example division of the parameters we adopted.}

\begin{tcolorbox}[colback=gray!20, colframe=gray!20, width=\columnwidth]
\textbf{Answer to RQ1:} Diff portion and diff thresh have minimal influence on the algorithm; Smaller \textit{vt}, shallower depth, and larger \textit{ws} produce fewer but larger and more accurately segmented areas; Larger \textit{vt}, greater depth, and smaller \textit{ws} generate smaller and finer divisions, albeit with reduced accuracy.
\end{tcolorbox}

\subsection{RQ2: How does \methodname compare to various methods?}

This RQ assesses \methodname's performance in design-to-code generation by comparing it against various prompting strategies across multiple automatic metrics.

\subsubsection{High-level Performance} Table~\ref{tab:gpt-4o-overall} presents the high-level performance comparison of DCGen and several baseline methods across the CLIP and Code metrics. DCGen outperforms all other methods, \add{achieving a notable 85.62\% on CLIP and 15.34\% on Code. Specifically, it surpasses the self-refine approach by 1.66\% on CLIP and 1.69\% on Code, highlighting its superior design-to-code generation capabilities. Notably, the Chain-of-Thought (CoT) method, while improving the text generation (Code: 13.64\%)}, does not significantly outperform other methods in visual fidelity.
 
\subsubsection{Fine-grained Metrics} 
Table~\ref{tab:fine-grain} showcases the fine-grained performance comparison of the methods in terms of Block-Match, Text, Position, and Color metrics. DCGen excels in these areas, achieving the highest scores \add{for Block-Match (81.7\%), Text (97.8\%), Position (83.6\%), and Color (85.4\%). The results suggest that DCGen provides a more accurate and complete design-to-code transformation, with particularly strong improvements in color accuracy (3.5\%) and text generation (1.6\%)}.

\begin{tcolorbox}[colback=gray!20, colframe=gray!20, width=\columnwidth]
\textbf{Answer to RQ2:} \methodname consistently outperforms other design-to-code techniques on both high-level and fine-grained metrics, due to the focused image segments.
\end{tcolorbox}

\begin{table}[t]
\vspace{-0.1in}
        \footnotesize
    \begin{minipage}{0.25\textwidth}
        \centering

        \caption{\add{\methodname's performance on GPT-4o (\%).}}
        \label{tab:gpt-4o-overall}
        \begin{tabular}{@{}lcc@{}}
        \toprule
        Method & CLIP & Code \\ \midrule
        Direct                   & {83.72}        & {12.76}        \\
        CoT                      & {83.64}        & {13.64}        \\
        Self Ref.              & {83.96}        & {13.65}        \\ 
        \methodname             & \textbf{ {85.62}}        & \textbf{ {15.34}} \\
        \midrule
        {Pix2Code}                 & {73.35}        & {9.70} \\
        \bottomrule
        \end{tabular}

    \end{minipage}
    \begin{minipage}{0.45\textwidth}
       \centering
       % \scriptsize
        \caption{\add{Fine-grained comparison of different methods (\%). }}
        \label{tab:fine-grain}
        \begin{tabular}{lcccc}
        \toprule
        Method       & Block-Match & Text   & Position & Color   \\
        \midrule
        Direct       & 76.1        & 95.9   & 81.6     & 81.8    \\
        CoT          & 79.5        & 88.4   & 76.9     & 76.2    \\
        Self Ref.  & 81.4        & 96.1   & 82.7     & 81.9    \\
        DCGen        & \textbf{81.7}        & \textbf{97.8}   & \textbf{83.6}     & \textbf{85.4}    \\
        \bottomrule
        \end{tabular}
    \end{minipage}
    \begin{minipage}{{0.25\textwidth}}
        \centering
   % \scriptsize
        \caption{\add{DCGen + GPT-4o on Design2Code-Hard (\%).}}
        \label{tab:d2c_hard}
        \begin{tabular}{@{}lcc@{}}
        \toprule
        Method & CLIP & Code \\ \midrule
        Direct                   & 81.04        & 14.37       \\
        CoT                      & 79.97        & 14.90       \\
        Self Ref.              & 80.43        & 15.20       \\ 
        DCGen             & \textbf{ {84.02}}        & \textbf{ {19.48}} \\
        \midrule
        {Pix2Code}                 & {73.56}        & {11.17} \\
        \bottomrule
        \end{tabular}
    \end{minipage}
\end{table}

\begin{table}[t]

\vspace{-15pt}
\end{table}

\subsection{RQ3: How robust is \methodname for complex designs?}
 \add{To evaluate the limits of DCGen, we test it on the Design-to-Code-Hard dataset, which consists of 80 highly complex webpages from GitHub Pages. This dataset presents significant challenges for state-of-the-art MLLMs in converting visual designs into code~\cite{Si2024Design2CodeHF}. Despite the increased difficulty, DCGen achieves the best performance in both code and visual similarity, as shown in Table~\ref{tab:d2c_hard}.}

\add{To further understand DCGen's role in improving results on complex webpages, we quantify website complexity using multiple metrics: image size, DOM depth, and the number of HTML tags, and focus on visual metrics (CLIP score) and code-generation metrics. We further include the Design-to-Code-Hard dataset, comprising 80 particularly challenging webpages, alongside our original dataset to provide a comprehensive assessment. The results measured by DOM depth and the number of HTML tags are presented on our website, which shows a moderate improvement of around 2\% across different complexities.} \add{The results measured by image size are summarized in Table~\ref{tab:img_size}:
\begin{enumerate}
    \item The separation rate remains high ($\ge$50\%) across various complexity levels.
    \item IoU scores are relatively stable (30\%-40\%) across most complexity scenarios, with a notable dip in the image size interval (15, 20].
    \item When measuring complexity by image size, DCGen consistently outperforms existing prompting methods, particularly for large images (15M-25M pixels), despite the IoU dip observed.
\end{enumerate}}

\add{These findings confirm that DCGen primarily benefits the conversion of large design images into code by breaking complex screenshots into smaller, more manageable visual segments. Even with imperfect segmentation, this strategy helps MLLMs better concentrate on detailed sub-tasks, significantly reducing omissions, distortions, and misarrangements. This finding is consistent with our motivating study, which revealed that decomposing large screenshots into smaller visual components improves model performance by facilitating focused reasoning steps.}

\begin{tcolorbox}[colback=gray!20, colframe=gray!20, width=\columnwidth]
\textbf{Answer to RQ3:} \methodname is robust to complex websites.\add{ It primarily benefits the conversion of large design images into code by breaking complex screenshots into smaller, more manageable visual segments.}
\end{tcolorbox}

\begin{table}
\caption{\add{Performance comparison across different input image sizes (in million pixels). All metrics are scaled by 100. Prompt$_{max}$ denotes the maximum score among self-refine, CoT, and direct prompting. ``DCGen Advantage'' indicates performance improvement of DCGen over Prompt$_{max}$, with positive values highlighted in green and advantages greater than 5 underlined.}}
\label{tab:img_size}
\centering
\footnotesize
\begin{tabular}{lccccc}
\toprule
\textbf{Image size (M)} & (0, 5] & (5, 10] & (10, 15] & (15, 20] & (20, 25] \\
\midrule
Sep Rate               & 58.8 & 70.8 & 75.0 & 84.1 & 51.4 \\
IoU                    & 43.6 & 41.5 & 36.5 & 21.2 & 45.1 \\
\midrule
CLIP Prompt$_{max}$            & 84.7 & 83.4 & 76.1 & 72.9 & 75.2 \\
CLIP DCGen             & 86.3 & 83.9 & 80.2 & 81.3 & 82.7 \\
DCGen Advantage        & \advColor{1.6} & \advColor{0.5} & \advColor{4.1} & \advColor{8.4} & \advColor{7.5} \\
\midrule
Code Prompt$_{max}$            & 15.9 & 11.3 & 5.5  & 11.5 & 4.5  \\
Code DCGen             & 18.2 & 15.7 & 5.9  & 27.1 & 6.5  \\
DCGen Advantage        & \advColor{2.3} & \advColor{4.4} & \advColor{0.4} & \advColor{15.6} & \advColor{2.0} \\
\bottomrule
\end{tabular}
\end{table}

\subsection{RQ4: How generalizable is \methodname for other SOTA MLLMs?}
In this RQ, we examine the generalizability of \methodname by employing the methodology on different MLLMs as backbones. Tables~\ref{tab:gemini-methods} and~\ref{tab:gemini2-methods} display the overall performance of \methodname on Gemini-1.5-flash and Gemini-2.0-flash. Our results indicate that \methodname is highly adaptable to the Gemini-1.5-flash model, achieving notable gains of 3.9\% in visual level metrics. \add{For Gemini-2.0, \methodname enhances the visual similarity between the original and the generated websites by 0.9\%, outperforming competing methods.} For Claude-3-Sonnet, \methodname also enhances the visual similarity by 3.8\%, outperforming competing methods. As for code similarity, we observe that self-refine prompting can help Claude-3 preserve text content in the screenshots (e.g., paragraphs and headings), leading to a performance jump for code similarity under the self-refine prompt. Overall, the experiment demonstrates the generalizability of \methodname on various MLLMs, showing the superiority of the divide-and-conquer strategy on UI Code generation.

\begin{tcolorbox}[colback=gray!20, colframe=gray!20, width=\columnwidth]
\textbf{Answer to RQ4:} \methodname is effective across different MLLMs in generating UI code from design, which demonstrates the generalizability of the proposed divide-and-conquer strategy.
\end{tcolorbox}

\begin{table}[htbp]
\small
    \centering
    \begin{minipage}{.32\linewidth}
        \centering
        \caption{Gemini-1.5-flash.}
        \label{tab:gemini-methods}
  
        \begin{tabular}{@{}lcc@{}}
            \toprule
            Gemini & CLIP & Code \\
            \midrule
            Direct & 72.0 & 16.6  \\
            CoT  & 72.1 & 16.8  \\
            Self-Refine & 73.7 & 16.1  \\
            \methodname & \textbf{77.6} & \textbf{17.8} \\
            \bottomrule
        \end{tabular}
    
    \end{minipage}%
    \begin{minipage}{.32\linewidth}
        \centering
        \caption{\add{Gemini-2.0-flash.}}
        \label{tab:gemini2-methods}
    
        \begin{tabular}{@{}lcc@{}}
            \toprule
            Gemini & CLIP & Code \\ 
            \midrule
            Direct & 86.7 & 17.2 \\
            CoT & 86.4 & 17.5 \\
            Self-Refine & 86.5 & 17.5  \\
            \methodname & \textbf{87.4} & \textbf{18.4} \\
            \bottomrule
        \end{tabular}
     
    \end{minipage}
    \begin{minipage}{.32\linewidth}
    \centering
       \caption{Claude-3-Sonnet.}
        \label{tab:claude-methods}

        \begin{tabular}{@{}lcc@{}}
            \toprule
            Claude-3 & CLIP & Code \\ 
            \midrule
            Direct & 73.8 & 13.5 \\
            CoT & 73.2 & 13.3 \\
            Self-Refine & 77.9 & \textbf{16.5}  \\
            \methodname & \textbf{81.7} & 13.5 \\
                    \bottomrule
        \end{tabular}
        % \centering
        % \caption{DCGen’s CLIP on InternVL2.}
        % \label{tab:internvl-methods}
        % \vspace{-10pt}
        % \begin{tabular}{@{}lccc@{}}
        %     \toprule
        %     Internvl2 & 8B & 40B & Gain \\ 
        %     \midrule
        %     Direct & 71.6 & 78.9 & +7.3 \\
        %     CoT & 68.4 & 68.7 & +0.3\\
        %     Self-Refine & \textbf{71.6} & \textbf{78.5} & +6.9  \\
        %     \methodname & 68.7 & 77.2 & +8.5 \\ 
        %     \bottomrule
        % \end{tabular}
        % \vspace{-10pt}
    \end{minipage}
    \vspace{-5pt}
\end{table}

\subsection{RQ5: What is the human's feedback on \methodname in real-world application development workflows?}
While automatic metrics offer a detailed breakdown of model performance, it is crucial to consider human feedback to validate the effectiveness of DCGen. To this end, we conduct a series of \textit{human evaluations} to compare the performance of different methods and directly assess the practical utility of DCGen in real-world application development workflows.

\subsubsection{Pairwise Method Comparison}
In this study, we recruit five PhD students from our university as annotators. Following the conventional practice of evaluating instruction following LLMs~\cite{Zhou2023LIMALI, Dubois2023AlpacaFarmAS}, we ask the annotators to compare pairs of generated webpages (one from the baseline, the other from the tested methods) to decide which one is more similar to the reference. We use GPT-4o Direct Prompting as the baseline and collect the other three methods’ Win/Tie/Lose rates against this baseline. Each pair will count as Win (Lose) only when Win (Lose) receives the majority vote ($\ge 3$). All other cases are considered Tie.

\begin{figure}[ht]
    \centering
    \includegraphics[width=0.66\linewidth]{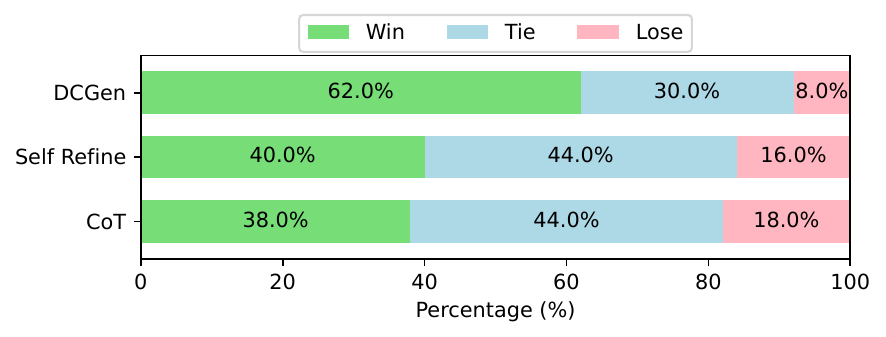}
    \vspace{-10pt}
    \caption{Human pairwise preference evaluation results with GPT-4o Direct Prompting as
the baseline. A higher win rate and lower lose rate suggest better quality as judged by human annotators.}
    \label{fig:pairwise}
% \vspace{-15pt}
\end{figure}

\textbf{Results.} We conducted pairwise comparisons between \methodname and baseline methods, with the results shown in Figure~\ref{fig:pairwise}. \methodname outperformed direct prompting in 62\% of cases and underperformed in just 8\%, showing its effectiveness in generating UI code.
Meanwhile, we find that \methodname is substantially better than other baselines, while both self-revision and CoT prompting show marginal improvements over direct prompting (winning in 38-40\% cases and losing in 16-18\% cases). These findings suggest that automatic metrics align with human perception, demonstrating that \methodname can produce more visually similar target webpages than competing methods under human evaluation.

\begin{figure}[ht]
    \centering
    \begin{subfigure}[b]{0.32\columnwidth}
        \centering
        \includegraphics[width=\textwidth]{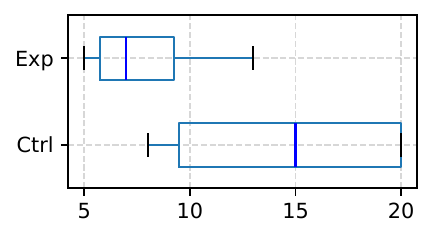}
        \caption{Time distribution (minutes).}
        \label{fig:usefulness-time}
    \end{subfigure}
        \begin{subfigure}[b]{0.32\columnwidth}
        \centering
        \includegraphics[width=\textwidth]{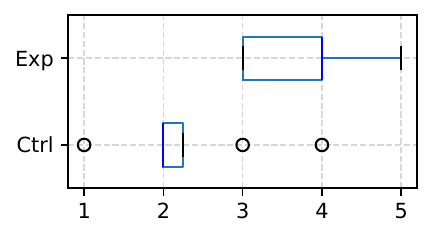}
        \caption{Similarity distribution.}
        \label{fig:usefulness-sim}
    \end{subfigure}
    \begin{subfigure}[b]{0.3\columnwidth}
            \centering
            \small
            \begin{tabular}{lcc}
            \toprule
             Measures    & Ctrl & Exp \\
            \midrule
            Time (min)  & 14.5    & \textbf{8} \\
            Similarity  & 2.25    & \textbf{3.75}  \\
            \bottomrule
            \vspace{3pt}
            \end{tabular}
            \caption{Average result.}
            \label{fig:usefulness-avg}
    \end{subfigure}
        \vspace{-8pt}
    \caption{The comparison of the experiment (Exp) and control group (Ctrl) on time and similarity.}
    \label{fig:usefulness}
\end{figure}

\subsubsection{Usefulness Evaluation}
While the pairwise evaluation offers a visual comparison among methods, we further assess the practicality of \methodname in accelerating the development workflow. In specific, we implement \methodname into a user-friendly demo tool (detailed in Section \ref{sec:discussion}) and recruit four PhD students and research staff from our institution to participate in our study, following the methodology from Chen et al.~\cite{Chen2018FromUI}. Each participant is tasked with converting two website UI images into corresponding UI code. The two images are selected to represent different levels of complexity: one simple and one more complex. Participants are randomly assigned to one of two groups: the experimental group (P1, P2), which uses DCGen to implement the UI code, and the control group (P3, P4), which is allowed to use any preferred tool (e.g., GPT-4o). A pre-study background survey confirms that all participants have similar development experience, being familiar with HTML and CSS and having previously developed one to two web applications for their work. Each participant has up to 20 minutes to complete each design.

We record the time each participant takes to implement the UI designs and capture screen recordings of their development processes, which are available in our supplementary materials. After the experiment, two research staff, who are not involved in the study, evaluate the similarity of the implemented GUIs to the original UI images using a five-point Likert scale~\cite{joshi2015likert} (1: not similar at all, 5: almost identical). The judges are blinded to which group produces each GUI.

\textbf{Results.}
Figure \ref{fig:usefulness} highlights UI implementation results from the experiment group (w/ DCGen) and control groups (w/o DCGen). First, the box plot in Figure \ref{fig:usefulness-time} shows that the experiment group (Exp) completes the design-to-code task twice as fast as the group without DCGen. The control (Ctrl) group also shows greater variability, with some participants taking up to 20 minutes. The average time of the control group is underestimated because one participant failed to complete a UI image within 20 minutes. In contrast, all participants in the experiment group finish all the tasks within 15 minutes. 
Second, as shown in Figure \ref{fig:usefulness-sim}, the experiment group consistently achieves higher similarity scores compared to the control group. The Pearson Correlation Coefficient between the similarity scores given by the two judges in Figure\ref{fig:usefulness-sim} is 0.71, suggesting a strong agreement between the judges. Finally, Figure \ref{fig:usefulness-avg} summarizes these results, confirming that the experiment group using DCGen has lower completion times and higher similarity scores on average, indicating our tool can help the experimental group implement UI images faster and more accurately.

\begin{tcolorbox}[colback=gray!20, colframe=gray!20, width=\columnwidth]
\textbf{Answer to RQ5:} Human evaluations demonstrate
that DCGen can produce more visually similar target webpages than competing methods, and the DCGen tool can significantly help developers implement webpages faster and more similar to the UI designs.
\end{tcolorbox}

\section{Discussion}\label{sec:discussion}
Based on the users' experience of interacting with \methodname, we discuss how it could contribute to transferring UI designs to code for developers, as well as identify potential limitations of \methodname for future improvements.

\begin{figure}
\begin{subfigure}[b]{0.68\linewidth}
     \centering
    \includegraphics[width=\linewidth]{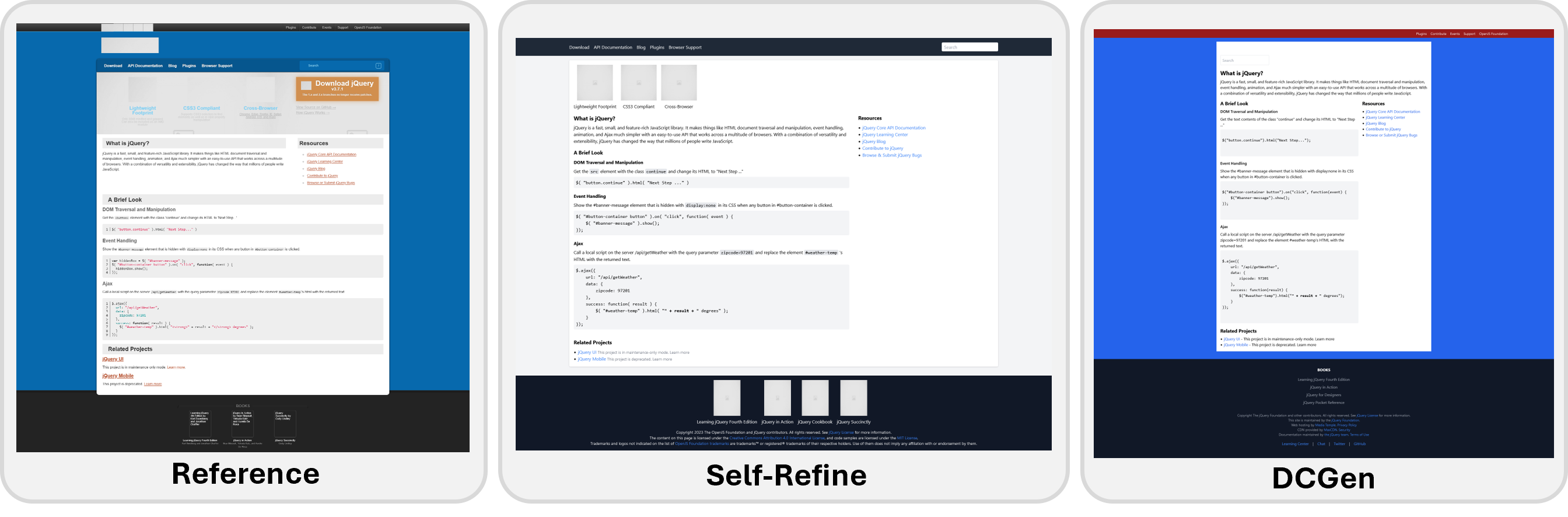}
\end{subfigure}
\begin{subfigure}[b]{0.7\linewidth}
    \centering
    \includegraphics[width=\linewidth]{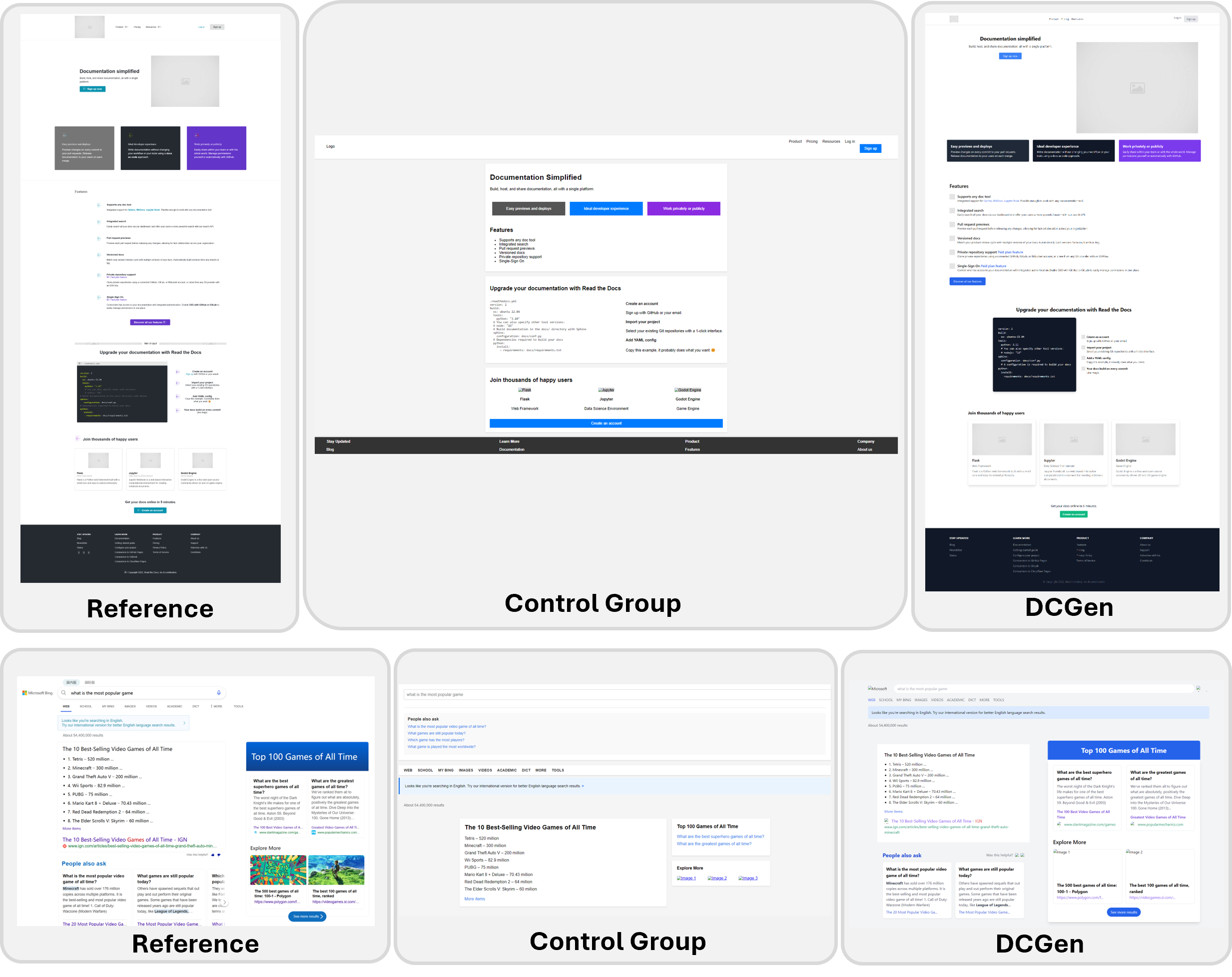}
\end{subfigure}
\caption{Success cases. The first row shows an example from RQ2, and the rest shows examples from the usefulness study (RQ5).}
    \label{fig:success_case}
        \vspace{-15pt}
\end{figure}

\textbf{Success stories.}
The first row of Figure~\ref{fig:success_case} illustrates a success case from Section 7 RQ2, which shows that compared to the self-refine method, DCGen restores the color and layout of the webpage, and removes nonexistent pictures from the webpage thus reducing distortion. The rest of Figure~\ref{fig:success_case} provides examples from the usefulness study (RQ5). DCGen demonstrates its effectiveness by significantly reducing element omissions and distortions, leading to faster development and improved webpage quality.

In the usefulness study (RQ5), we observe that both developers in the control group follow a similar iterative refinement pattern to generate UI code, even though they are unaware of DCGen's methodology. Specifically, they start by prompting the model to produce an initial webpage, then iteratively refine the unsatisfactory parts by providing more detailed prompts with a screenshot of the specific areas. Such manual narrowing of task scope aligns with DCGen's automated divide-and-conquer approach.

\begin{figure}[t]
    \centering
        \centering
        \includegraphics[width=0.53\textwidth]{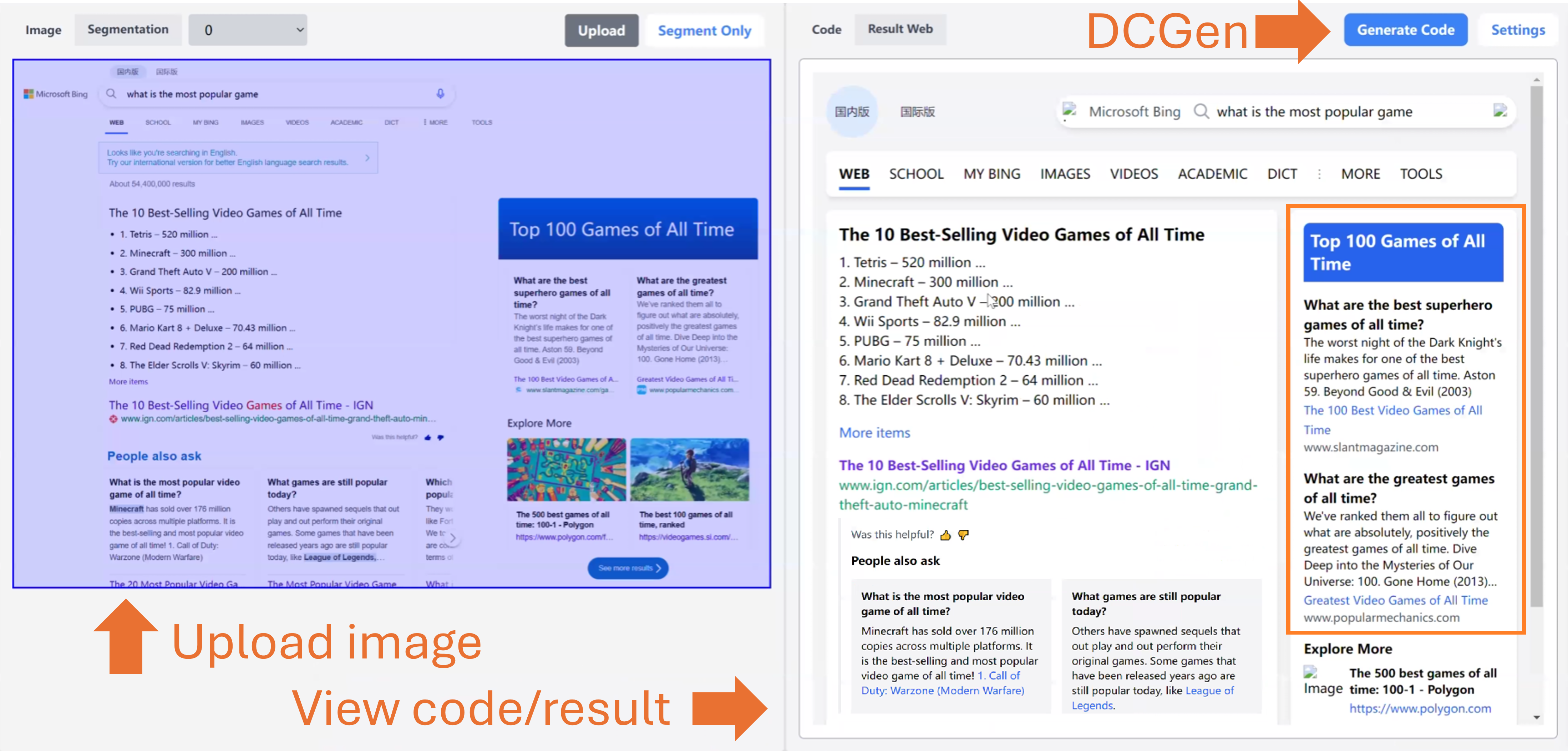}
        \caption{Generate code for the entire webpage.}
        \label{fig:dcgenui}
    \vspace{-15pt}
\end{figure}

\textbf{Tool features.}
Figure~\ref{fig:dcgenui} shows a screenshot of DCGen's user interface as a development tool, with the design image on the left and the code and rendered webpage on the right. DCGen enables developers to generate an entire webpage, inspect the code for individual image segments, and regenerate code for any specific segment. By automating the divide-and-conquer process, DCGen eliminates the need for detailed, repetitive prompts, speeding up development and enabling the creation of higher-quality webpages in less time compared to the control group.

% \begin{figure}[t]
%     \centering
%     \begin{subfigure}[b]{0.48\columnwidth}
%         \centering
%         \includegraphics[width=\textwidth]{Sections/figs/dcgenui1.png}
%         \caption{Generate code for the entire webpage.}
%         \label{fig:dcgenui1}
%     \end{subfigure}
%     \hspace{3px}
%         \begin{subfigure}[b]{0.48\columnwidth}
%         \centering
%         \includegraphics[width=\textwidth]{Sections/figs/dcgenui2.png}
%         \caption{View the code of an image segment.}
%         \label{fig:dcgenui2}
%     \end{subfigure}
%         \begin{subfigure}[b]{0.48\columnwidth}
%         \centering
%         \includegraphics[width=\textwidth]{Sections/figs/dcgenui3.png}
%         \caption{Generate code for an image segment.}
%         \label{fig:dcgenui3}
%     \end{subfigure}
%         % \vspace{-20pt}
%     \caption{DCGen user interface and functionalities.}
%     \label{fig:dcgenui}
%     \vspace{-15pt}
% \end{figure}

\textbf{Failure analysis.} While DCGen effectively reduces element omission and distortion, its improvement in addressing element misarrangement is limited. A failure case is illustrated in Figure~\ref{fig:dcgenui}, within the orange rectangular area. In the original image, the paragraphs are horizontally arranged, but the generated paragraphs are vertically aligned. Upon further investigation, we found that this error occurs during the assembly process, as the leaf solver LLM correctly generates horizontally arranged paragraphs for this area, while the assembly LLM mistakenly changes the paragraphs to a vertical layout. We attribute the issue to the inherent visual limitations of MLLMs, as identified by Tong et al.~\cite{Tong2024EyesWS}, which suggest that MLLMs still face challenges in recognizing certain simple visual patterns, such as positional arrangements.

However, since DCGen allows users to review and regenerate code for any specific image segment, the erroneous code segment can be easily replaced with the correct version in the final webpage. Thus, this failure case has minimal impact on DCGen's practical effectiveness.

\section{Threat to Validity}
% We have identified the following threats to the validity of our work

\textit{The selection of backbone models.}
This study employs four popular instruction-following and multimodal LLMs to demonstrate the effectiveness of \methodname.
While other LLMs might be adapted for our methodology, we found that smaller parameter models fail to perform complex tasks such as the assembly process.
Our future work involves extending our implementation to other emerging models, showcasing \methodname's generalization abilities.
% Another threat is the \textbf{high demand for model capabilities} in instruction understanding, reasoning, and code generation. Therefore, our approach may not perform well with less powerful models. 

\textit{Robustness of set thresholds and chosen separation strategy.}
Assumptions regarding the segmentation strategy and manually set thresholds may limit DCGen's applicability in more complex designs. To address this concern, we conduct RQ2 to evaluate DCGen's robustness to complex designs. The results show that DCGen delivers superior performance across various website complexities, indicating its robustness in handling both simple and intricate webpages.

\section{Related Work}

\subsection{Image-to-code Generation}
The most informative AI techniques used for generating code from images can be classified into three broad categories: (i) Convolutional Neural Networks (CNNs)~\cite{Asiroglu2019AutomaticHC, Beltramelli2018Pix2c, Xu2021Image2e, Moran2018MachineLP, Chen2022CodeGF, Cizotto2023WebPF, Chen2018FromUI}, (ii) Computer Vision (CV) and Optical Character Recognition (OCR)~\cite{Nguyen2015ReverseEM, Natarajan2018P2a}, and (iii) other deep learning models~\cite{Wu2021ScreenP, Abdelhamid2020DeepLP}. 
The work \cite{Beltramelli2018Pix2c} utilizes CNNs and LSTM to extract features from GUI images to generate a domain-specific language (DSL).
However, DSLs are not widely used in practical development, which inconveniences the industry's UI developers~\cite{Xu2021Image2e}. An alternative approach~\cite{Chen2018FromUI} using CNNs to extract visual features from UI images is more widely applicable than generating fixed DSL code.
Based on CNNs, Andre et al.~\cite{Cizotto2023WebPF} pioneered using Class Activation Mapping (CAM), further improving the performance and interpretability of the image-to-code generation task. More recently, Interaction2Code~\cite{Xiao2024Interaction2CodeHF} explored interactive web code generation from design images, and MRWeb~\cite{Wan2024MRWebAE} investigated the generation of multi-page resource-aware web code from UI designs, which extends page-level image-to-code generation to project-level.

\subsection{Multimodal Language Models}
Multimodal Language Models (MLLM) refer to LLM-based models capable of receiving, reasoning, and outputting multimodal information~\cite{Yin2023ASO}. In recent years, the trend of using LLMs as decoders in vision-language tasks has gained significant traction. Pioneering studies such as VisualGPT \cite{chen2022visualgpt} and Frozen \cite{tsimpoukelli2021multimodal} have demonstrated the advantages of employing a pre-trained language model as a decoder. Following this, Flamingo \cite{alayrac2022flamingo} was developed to align a pre-trained vision encoder and language model using gated cross-attention. Furthermore, BLIP-2 \cite{li2023blip} introduced the use of Flan-T5 \cite{chung2024scaling} with a Q-Former to efficiently align visual features with the language model. PaLM-E \cite{driess2023palm}, featuring 562 billion parameters, was developed to integrate real-world continuous sensor modalities into an LLM. More recently, GPT-4o \cite{openai_gpt4o} and Gemini-2.5~\cite{google_gemini_api} demonstrate enhanced visual understanding and reasoning abilities following pre-training on a vast collection of aligned image-text data.
% Prior to MLLM, extensive research focused on multimodal paradigms, which can be broadly classified into discriminative~\cite{Radford2021LearningTV, Li2021AlignBF, Chen2019UNITERUI} and generative~\cite{Wang2022OFAUA, Cho2021UnifyingVT, Wang2021SimVLMSV} approaches. MLLMs fall under the generative category due to their sequence operation capabilities but distinguish themselves with two significant traits compared to traditional models: (1) They are based on LLMs with billion-scale parameters, which were previously unavailable, and (2) They utilize novel training paradigms to fully harness their potential, such as multimodal instruction tuning~\cite{Wei2021FinetunedLM, Liu2023VisualIT}, enabling the model to adapt to new instructions effectively. These advancements endow MLLMs with unique capabilities, such as generating website code from images~\cite{Zhu2023MiniGPT4EV}, comprehending the nuanced meanings of memes~\cite{Yang2023MMREACTPC}, and performing OCR-free mathematical reasoning~\cite{Driess2023PaLMEAE}.

\section{Conclusion}
% In this paper, we design and implement \methodname, a novel divide-and-conquer-based framework for effectively generating UI code from web screenshots. We evaluate \methodname across three SOTA MLLMs and demonstrate that it consistently outperforms various prompting strategies.

In this paper, we first present a motivating study that identifies prevalent failures in MLLMs during the design-to-code generation process. Drawing on these insights, we design and implement \methodname, a novel divide-and-conquer-based framework for effectively generating UI code from web screenshots. We evaluate \methodname across three SOTA MLLMs and demonstrate that it consistently outperforms various prompting strategies. Furthermore, \methodname shows robustness against variations in website complexity and is generalizable across different MLLMs, demonstrating its superiority in generating UI code.

\section*{Data Availability}
All the code, data, and user-friendly tool for DCGen are available at \url{https://github.com/WebPAI/DCGen} for replication and future research.

\section*{Acknowledgment}
This research was supported by the Singapore Ministry of Education (MOE) Academic Research Fund (AcRF) Tier 1 grant and Research Grants Council of the Hong Kong Special Administrative Region, China (No. SRFS2425-4S03 of the Senior Research Fellow Scheme).

% \section{Data Availability}
% All the code, data, and results have been released~\footnote{\url{https://drive.google.com/drive/folders/1FnR6MTKCSWFsUP__qO-J5YRhSB7RRDI-?usp=sharing}} for reproduction and future research.
\bibliographystyle{ACM-Reference-Format}
\bibliography{reference}


\end{document}